\documentclass[12pt,a4paper,oneside,onecolumn]{article}
\usepackage[cp1252]{inputenc}
\usepackage[english]{babel}
\usepackage{amsmath}
\usepackage{amssymb}
\usepackage{graphicx}
\usepackage{setspace}
\usepackage[a4paper,left=3cm,right=2cm,top=2cm,bottom=2cm]{geometry}
\usepackage[square,comma,numbers,sort&compress]{natbib}
\usepackage{sectsty}
\usepackage[symbol]{footmisc}
\allsectionsfont{\large\centering}
\makeatletter
\def\@seccntformat#1{\csname the#1\endcsname.\quad}
\makeatother
\begin{document}
\begin{flushleft}
\textit{\large Astronomy Reports, 2010, Vol. 54, No. 4, pp. 338---354.}
\footnote{Original Russian Text published in Astronomicheskii Zhurnal, 2010, Vol. 87, No. 4, pp. 379---396.}
\end{flushleft}
\vspace{\baselineskip}
\begin{center}
\begin{Large}
\textbf{The Appearance of a Radio-Pulsar Magnetosphere from a Vacuum with a Strong Magnetic Field.\linebreak Motion of Charged Particles}
\end{Large}
\\ \bigskip
\textrm{\large Ya. N. Istomin$^1$ and D. N. Sob'yanin$^2$}
\\ \bigskip
\textit{$^1$Lebedev Physical Institute, Russian Academy of Sciences,\\
53 Leninskii pr., Moscow, Russia}
\\ \medskip
\textit{$^2$Moscow Institute of Physics and Technology (State University),\\
Dolgoprudnyi, Moscow oblast, Russia}
\\ \bigskip
Received June 30, 2009; in final form, October 19, 2009
\\ \bigskip
\begin{large}
\textbf{Abstract}
\end{large}
\end{center}
\vspace{-0.3cm}
\par\textrm{The motion of electrons and positrons in the vacuum magnetosphere of a neutron star with a surface magnetic field of $B\sim10^{12}$~G is considered. Particles created in the magnetosphere or falling into it from outside are virtually instantaneously accelerated to Lorentz factors $\gamma\sim10^8$. After crossing the force-free surface, where the projection of the electric field onto the magnetic field vanishes, a particle begins to undergo ultra-relativistic oscillations. The particle experiences a regular drift along the force-free surface simultaneous with this oscillatory motion.}
\newpage
\section{INTRODUCTION}
The stationary structure of the magnetosphere of a radio pulsar has been studied in considerable detail. Here, this refers not to the vacuum magnetosphere, but instead to a magnetosphere filled with dense electron-positron plasma. This is associated with the fact that the radio emission generated in the magnetosphere by the flux of charged particles requires the action of some mechanism producing a dense plasma. In one such mechanism, proposed by Sturrock \citep{Sturrock1971} and substantially developed by Ruderman and Sutherland \citep{RudermanSutherland1975}, the efficient creation of electron-positron pairs by gamma-rays with energies exceeding twice the electron rest-mass energy is possible in the strong magnetic field at the surface of the neutron star---radio pulsar, $B\simeq 10^{12}$~G \citep{Klepikov1954,Erber1966}. In turn, energetic gamma-rays are emitted by electrons and positrons during their motion in the magnetosphere along magnetic-field lines having appreciable curvature. Such photons are called curvature photons. The emission and absorption of photons in a magnetic field accompanied by the creation of electron-positron pairs provides a theoretical basis for the steady-state creation of plasma in the magnetosphere of a magnetized, rotating neutron star. The density of the forming plasma $n$ appreciably exceeds the so-called Goldreich-Julian density \citep{GoldreichJulian1969},
$n_{GJ}=|\mathbf{\Omega}\cdot\mathbf{B}|/2\pi ce$,  which provides stationary rotation of the magnetosphere right out to the light cylinder, $R_L=c/\Omega$. Here, $\Omega$ is the angular velocity of the star’s rotation, $c$ is the speed of light, and $e$ is the positron charge. The multiplicity parameter for the creation of the electron-positron plasma $\lambda=n/n_{GJ}$ is large: $\lambda\simeq 10^4{-}10^5$.

However, it is difficult to understand from observations of stationary radio pulsars what radio-emission mechanisms are operating in their magnetospheres, and where and how the plasma is produced. Tracing the dynamics of the development of the emission at different frequencies could be very important for our understanding of the physical processes occurring in the magnetospheres of radio pulsars \citep{GurevichIstomin2007}. Moreover, many observations of non-stationary radio pulsars have recently appeared. These are first and foremost so-called switching radio pulsars, from which radio emission is observed only during certain time intervals appreciably exceeding the rotational period of the star. For example, PSR B1931+24 is ``on'' for 5--10 days, then ``off'' for 20--25 days \citep{KramerEtal2006}; PSR J1832+0029 is ``on'' for about 300 days and then ``off'' for about 700 days (see, e.g., the review~\citep{Kramer2008}). Measurements indicate that the deceleration of the  rotation, i.e., the loss of energy, is appreciably different during the ``on'' and ``off'' periods. Since the power of the radio emission is a negligible fraction of the total rotational energy lost by the neutron star, it is natural to suppose that quiet periods are associated with the cessation of the generation of plasma in the magnetosphere. We can then distinguish loss mechanisms associated with the radiation of magnetodipole waves in vacuum (``off'' periods) and with the emission of the pulsar wind in the form of a flux of electron–positron plasma flowing from the magnetosphere (``on'' periods).

In addition to switching pulsars, a group of so-called nulling pulsars has long been known, which likewise display no radio emission during certain intervals, but with these being not as regular as those for switching pulsars. Differences in the rotational deceleration have not yet been measured for nulling pulsars. For example, PSR B1944+17
does not have detectable radio emission 55\% of the time \citep{Ritchings1976}. Of the
23 pulsars studied in \citep{WangEtal2007}, 7 display nulling fractions exceeding 40\%, while this fraction reaches 95\% for PSR J1502--5653 and PSR J1717--4054.

Another group of non-stationary radio sources has recently been observed: rotating radio transients (RRATs), which are sporadically flaring radio sources. The phases are preserved during these flares, and the corresponding measured periods are characteristic of ordinary radio pulsars \citep{McLaughlinEtal2006}. There is no doubt that these are also rotating neutron stars. However, the nature of their activity is quite unclear,
as is testified to by the presence of numerous and varied models for these objects, such as models invoking precession \citep{ZhuXu2006}, reversal of the direction of the radio beams \citep{ZhangEtal2007}, re-activation of ``dead'' pulsars \citep{ZhangEtal2007}, interaction of the magnetosphere with a disk \citep{Li2006}, drift waves \citep{LomiashviliEtal2007}, and even such exotic objects as the remnants of quark novae \citep{OuyedEtal2008}. Naturally, such models are open for discussion. For example, when
considering re-activation of the radio emission of pulsars located to the right of the ``death line'' in the $P{-}\dot{P}$ diagram, we must know the exact position of
this line. Timing data for the RRAT J1819--1458 suggest that the magnetic field at the stellar surface is approximately $5\times10^{13}$~G, so that it exceeds the Schwinger field \citep{McLaughlinEtal2006,EsamdinEtal2008}. In such a strong magnetic field, it is necessary to consider the splitting of photons taking into account their polarization and different conditions for the creation of pairs than in the case of a weak magnetic field. As a consequence, the death line for pulsars with strong magnetic fields has a slope of $11/3$ \citep{IstominSobyanin2007,IstominSobyanin2008} rather than the usual slope of $11/4$, which occurs only in the case of weak magnetic fields.

In our view, all these sources exhibit a non-stationary generation of plasma in neutron-star magnetospheres. It is therefore important to understand how the magnetosphere of a rotating, magnetized neutron star is filled with plasma, which is the reason for the operation of radio pulsars. An important aspect of this is the state of the magnetosphere of the rotating neutron star before it becomes a radio pulsar, or after the source of plasma has switched off in its magnetosphere. If this source does not operate, the dense plasma falls onto the surface of the neutron star in a closed magnetosphere over the characteristic time $t\simeq R_L/c=\Omega^{-1}=P/2\pi$, which is less than the rotational period. In an open magnetosphere, the plasma flows out over the same time scale. This raises the question of what then remains in the magnetosphere.

One important discovery here was yielded by observations of the dynamics of the braking of the two switching pulsars PSR~B1931+24 and PSR~J1832+0029 \citep{KramerEtal2006,Kramer2008}, which showed that the deceleration of the star’s rotation continued after the disappearance of the radio emission. The rate of this deceleration was below the initial value by approximately a factor of $1.5$, but not by an order of magnitude. It is not possible to explain the observed braking if plasma with a density of the order of $n_{GJ}$ remained in the magnetosphere after the pulsar switched off, so that the screening of the longitudinal electric field continued. In this case, the magnetodipole radiation would be fully screened. If there were energy losses associated with the outflow of plasma with the density $n_{GJ}$ from an open magnetosphere, the particle energies would have to be of the order of $10^{11}$~eV, which would require the presence of an acceleration region in which the longitudinal electric
field was not equal to zero. An energetic production of plasma with multiplicity $\lambda\gg1$ would then begin in this region, in contradiction with the fact that the pulsar was switched off. Moreover, it may be that there is no plasma in an open magnetosphere, but that it occupies some closed region. However, as was shown in \citep{GurevichIstomin2007}, magnetodipole radiation will be weakened by a
factor of $(R/R_L)^{3/2}\sim10^{-6}{-}10^{-4}\ll1$ in this case, compared to the case of a pure vacuum, which is not observed. The only reasonable conclusion is that there is no plasma in the pulsar’s magnetosphere when it is not in its operational state, and that the observed rotational deceleration is brought about via magnetodipole radiation. In this case, the energy losses have the same order of magnitude as for the operational pulsar, when the generation of plasma and electric current in the magnetosphere lead to braking of the rotation. We do not discuss here what this would imply for various plasma-production models developed over many years; the absence of plasma in the switched-off state ($n\ll n_{GJ}$) represents direct evidence from available observational data, and not a new proposed model.

We are concerned here with an initial investigation into the ``ignition'' of the magnetospheres of neutron stars---the dynamics of the filling of the vacuum magnetosphere with electrons and positrons created in the magnetosphere. The second section considers the electromagnetic fields of the inner vacuum magnetosphere and describes the force-free surface where the electric field component along the magnetic field vanishes. The following sections are dedicated to the dynamics of the motion of charged particles near the force-free surface.

\section{STRUCTURE OF THE VACUUM MAGNETOSPHERE}

It will be convenient for us to consider the electromagnetic field around a neutron star in the spherical coordinates $(r,\theta,\varphi)$. We take the polar axis to coincide with the rotational axis of the star, which determines the direction of its angular-momentum vector $\mathbf{\Omega}$. Here, $r$ is the distance from the center of
the star to a given point, $\theta$ the polar angle measured from the rotational axis, and $\varphi$ the azimuthal angle. The electric and magnetic fields will be written in the form
\begin{equation*}
\begin{aligned}
\mathbf{E}&=E_r\mathbf{e}_r+E_\theta\mathbf{e}_\theta+E_\varphi\mathbf{e}_\varphi,\\
\mathbf{B}&=B_r\mathbf{e}_r+B_\theta\mathbf{e}_\theta+B_\varphi\mathbf{e}_\varphi,
\end{aligned}
\end{equation*}
where the unit vectors $\mathbf{e}_r, \mathbf{e}_\theta, \mathbf{e}_\varphi$ are mutually orthogonal and form a right-handed set. The electromagnetic field outside the neutron star was found by Deutsch \citep{Deutsch1955} and has the form

\noindent
magnetic field
\begin{equation}
\label{deutschMagneticField}
\begin{aligned}
B_r&=\frac{2m}{r^3}\,\Bigl[\cos\theta\cos\theta_m+\sin\theta\sin\theta_m\cos(\varphi-\varphi_m)\Bigr],\\
B_\theta&=\frac{m}{r^3}\,\Bigl[\sin\theta\cos\theta_m-\cos\theta\sin\theta_m\cos(\varphi-\varphi_m)\Bigr],\\
B_\varphi&=\frac{m}{r^3}\,\sin\theta_m\sin(\varphi-\varphi_m)
\end{aligned}
\end{equation}
and electric field
\begin{equation}
\label{deutschElectricField}
\begin{aligned}
E_r&=-k\frac{mR^2}{r^4}\,\left[\Bigl(\frac{3}{2}\cos2\theta+\frac{1}{2}\Bigr)\cos\theta_m+\frac{3}{2}\sin2\theta\sin\theta_m\cos(\varphi-\varphi_m)\right],\\
E_\theta&=-k\frac{mR^2}{r^4}\,\left[\sin2\theta\cos\theta_m+\Bigl(\frac{r^2}{R^2}-\cos2\theta\Bigr)\sin\theta_m\cos(\varphi-\varphi_m)\right],\\
E_\varphi&=k\frac{mR^2}{r^4}\,\Bigl(\frac{r^2}{R^2}-1\Bigr)\cos\theta\sin\theta_m\sin(\varphi-\varphi_m).
\end{aligned}
\end{equation}
Here, $(\theta_m,\varphi_m)$ are the polar and azimuthal angles of the magnetic axis, which is determined by the direction of the magnetic dipole moment $\mathbf{m}$, $k=\Omega/c$ is the wave number corresponding to the angular frequency of rotation of the neutron star $\Omega$, and $R$ is the radius of the star. The azimuthal angle $\varphi_m=\Omega t$ is chosen to that it is equal to zero at time $t=0$. Although we are primarily interested here in the case of an uncharged, rotating, magnetized neutron star possessing a high conductivity, the generalized solution of Deutsch for
a charged sphere can be found in \citep{Tiomno1973,Jackson1978}. A useful discussion of this question is also presented in the review \citep{MichelLi1999}. The solution of Deutsch refers to a dipolar magnetic-field distribution at the stellar surface. Its generalization to the case of an arbitrary, axially symmetrical magnetic-field distribution is presented in \citep{Soper1972}.

Formulas \eqref{deutschMagneticField} and \eqref{deutschElectricField} are valid until we consider the electromagnetic field at distances that are appreciably less than the radius of the light cylinder, $R_L=c/\Omega$. More precisely,
\begin{equation}
\label{quasistaticCondition}
(\Omega r_\perp/c)^2\ll1,
\end{equation}
where $r_\perp$ is the distance from the rotational axis of the neutron star to the point considered. As $r_\perp$ is increased in \eqref{deutschMagneticField} and \eqref{deutschElectricField}, terms of the expansion with even powers of $(\Omega r_\perp/c)$ begin to appear. Further, we will be interested in the generation of electron–positron plasma in the neutron-star magnetosphere. As we know, the single-photon creation of an electron–positron pair is efficient only in the presence of a fairly strong magnetic field. The magnetic field, which has a dipolar structure, rapidly falls off with distance $r$ from the center of the star ($\propto r^{-3}$). The creation of electron–positron pairs becomes inefficient at magnetic-field strengths of $\sim 10^{8}$~G. For pulsars with characteristic surface magnetic fields $10^{12}$~G, the characteristic distance at which the creation of pairs becomes possible is of the order of $(10{-}20)R$. Therefore, for typical stellar radii $R\simeq10$~km and rotational periods $P\sim0.1{-}1$~s, the parameter \eqref{quasistaticCondition} is of order $10^{-5}{-}10^{-3}$. Under these conditions, \eqref{deutschMagneticField} and \eqref{deutschElectricField}yield fairly precise results. An impression of the structure of the vacuum magnetosphere at distances comparable with $R_L$  can be obtained, for example, from \citep{FerrariTrussoni1973}.

To further analyze the dynamics of the particle motions in the vacuum magnetosphere of the neutron star, we will use the equation for the so-called force-free surface---the surface satisfying the equation $\mathbf{E}\cdot\mathbf{B}=0$. We are interested in this surface for the following reason. We must determine how the charged electrons and positrons that are created in the magnetosphere will move in the magnetosphere, and whether there will exist regions where they accumulate. We can immediately qualitatively assert (we will consider this question quantitatively below) that, after their creation, particles will virtually instantaneously be accelerated to relativistic speeds if the longitudinal electric field $E_\parallel=\mathbf{E}\cdot\mathbf{B}/B$ differs from zero. The only region where particles could collect is the force-free surface, at each point of which the longitudinal electric field is zero.

Using \eqref{deutschMagneticField} and \eqref{deutschElectricField} for the Deutsch electromagnetic field, it is straightforward to obtain an expression for the scalar product of the electric and magnetic vectors (see also \citep{Jackson1978}):
\begin{equation}
\label{EBdotProduct}
\mathbf{E}\cdot\mathbf{B}=-kr\frac{m^2}{r^6}\left[\Bigl(1-\frac{R^2}{r^2}\Bigr)
\cos\theta''\sin\theta_m+\frac{R^2}{r^2}4\cos^2\theta'\cos\theta\right].
\end{equation}
Here, we have introduced the angles $\theta'$ and $\theta''$ as follows:
\begin{equation*}
\label{thetaAngles}
\begin{aligned}
\cos\theta'&=\cos\theta\cos\theta_m+\sin\theta\sin\theta_m\cos(\varphi-\varphi_m),\\
\cos\theta''&=-\cos\theta\sin\theta_m+\sin\theta\cos\theta_m\cos(\varphi-\varphi_m).
\end{aligned}
\end{equation*}
Let us define these angles less formally. Let $\mathbf{e}_r=\mathbf{r}/r$ and
$\mathbf{e}_m=\mathbf{m}/m$ be unit vectors directed along the radius vector
$\mathbf{r}$ and along the magnetic axis $\mathbf{m}$. Further, we define the vector
$\mathbf{e}_n=\partial\mathbf{e}_m/\partial\theta_m$, lying in the plane of the vectors
$\mathbf{\Omega}$ and $\mathbf{m}$, directed orthogonal to $\mathbf{m}$ and indicating the direction of growth in $\theta_m$. Then, the angle $\theta'$ is the angle between the radius vector and the magnetic axis, and $\theta''$ is the angle between the radius vector and the $\mathbf{e}_n$ axis; i.e., $\cos\theta'=\mathbf{e}_r\cdot\mathbf{e}_m$ and $\cos\theta''=\mathbf{e}_r\cdot\mathbf{e}_n$.

Using \eqref{EBdotProduct}, we can easily obtain an equation for the force-free surface where $\mathbf{E}\cdot\mathbf{B}=0$:
\begin{equation}
\label{FFSequation}
r_{ffs}^2=R^2\left(1-4\frac{\cos\theta\cos^2\theta'}{\sin\theta_m\cos\theta''}\right).
\end{equation}

The structure of the force-free surface \eqref{FFSequation} is illustrated by the figure, which shows cross sections of the force-free surface by the plane $\varphi-\varphi_m=\{0,\pi\}$, passing through the rotational axis and magnetic axis,
and by the plane $\varphi-\varphi_m=\{\pi/2,3\pi/2\}$, passing through the rotational axis orthogonal to the previous plane, for various angles between the magnetic and rotational axes of the neutron star. An understanding of the form of the force-free surface can be obtained from \citep{FinkbeinerEtal1989}.

In the case of a co-axial rotator, the force-free surface is simply the equatorial plane, $\theta=\theta'=\pi/2$. For a rotator with an arbitrary inclination, the force-free
surface can be divided into two regions. One region has the form of two arched parts adjacent to the surface of the neutron star at the points of the equator and magnetic equator. The magnetic equator is the circle that is the cross section of the $\theta'=\pi/2$ plane passing through the center of the neutron star orthogonal to the magnetic axis, $\mathbf{m}$, and the stellar surface, $r=R$. The ordinary equator is the cross section of the $\theta=\pi/2$ plane passing through the center of the star orthogonal to the rotational axis, $\mathbf{\Omega}$, and the stellar surface. The force-free surface is adjacent to the surface of the neutron star at all points of the magnetic equator, which is not true at points of the ordinary equator.

The second region can be represented as a composition of two open sheets. One edge of each sheet is adjacent to the magnetic equator and the straight line passing through the center of the neutron star orthogonal to the rotational and magnetic axes. The other edge of the sheet extends to infinity, such that, for an arbitrary point lying on the force-free surface, $\theta''\rightarrow\pi/2$ as $r\rightarrow\infty$. At large $r$,
the sheet differs little from the $\theta''=\pi/2$ plane. However, the difference becomes appreciable at relatively small distances from the neutron star, and a cupola-like protrusion ending at the magnetic equator forms above the $\theta''=\pi/2$ plane. The two sheets are joined smoothly at the points of the line $\theta=\theta'=\theta''=\pi/2$, forming a single sheet. The points of this line cannot be described using \eqref{FFSequation}, because this equation is degenerate at these points, in connection with our choice of coordinates on the surface of the angular coordinates $(\theta,\varphi)$. We can directly verify that this line does indeed belong to the force-free surface using \eqref{EBdotProduct}.

As we can see from the figure, the dimensions of the closed arched parts of the force-free surface grow with $\theta_m$. When this angle reaches $\pi/2$, there is a break in the edges of the arches adjacent to the equator. These edges smoothly join with the edges of the open sheets that extend to infinity, forming an axially symmetrical, double-cupola figure specified by the equation $r^2=R^2(1+4\cos^2\theta')$. It is easy to see that the electric field is also orthogonal to the magnetic field when $\theta=\pi/2$. Thus, the force-free surface of an orthogonal rotator is a combination of the figure described above and the equatorial plane.

If the angle $\theta_m$ is increased further, the part of the force-free surface that was arch-like for $\theta_m<\pi/2$ is transformed into an open sheet when $\theta_m>\pi/2$, as a result of the edge breaking away from the equator and its transition to infinity. The open sheets, on the contrary, are transformed into arch-like parts of the surface, due to the attachment of their open edges to the equator. Note that we are speaking here of the breaking away of edges with some tentativeness, since it is more correct to speak of a reconnection of parts of the force-free surface as the angle $\theta_m$ passes through $\pi/2$. Indeed, if $\theta_m$ becomes equal to $\pi/2$, the closed arch-like parts become non-smooth while remaining continuous, and each of the arches is represented as a joining of the bulging part and a flat part belonging to the equatorial plane $\theta=\pi/2$ and lying inside the double-cupola structure noted above. This is also true of the open sheets, with the exception that their flat parts lie outside this structure. In the transition through $\theta_m=\pi/2$, there is a sort of reclosing of the parts of the force-free surface: the flat part of the open sheet becomes reconnected to the bulging part of the arch, and the flat part of the arch with the bulging part of the open sheet.

Taking this into account, we further consider everywhere only angles $\theta_m$ lying between $0$ and $\pi/2$. This does not limit the generality of our discussion, since a rotator with the angle $\chi>\pi/2$ between the vectors $\mathbf{m}$ and $\mathbf{\Omega}$ is equivalent to a rotator with $\theta_m=\pi-\chi<\pi/2$, with $\Omega$ replaced by $-\Omega$ in all formulas.

\section{MOTION OF CHARGED PARTICLES}

We will investigate the motions of particles in a vacuum magnetosphere using the classical Dirac–Lorentz equation
\begin{equation}
\label{DiracLorentzEquation}
m_e\ddot{x}^i=\frac{2}{3c^3}e^2\left[\dddot{x}^i+\frac{1}{c^2}\dot{x}^i\,\ddot{x}^k\ddot{x}_k\right]+F^i,
\end{equation}
where $x^i=(ct,\mathbf{r}^T)^T$ is a contravariant four-vector containing
the time $t$ and coordinates $\mathbf{r}=(x,y,z)^T$ of the particle in the laboratory frame ($T$ denotes transposition), $m_e$ the particle’s mass, $e$ the particle’s charge, and $c$ the speed of light. A dot above a four-vector denotes differentiation with respect to the proper time $\tau$ of the particle, i.e., the time in a frame comoving
with the particle. The differentials $dt$ of time in the laboratory frame and $d\tau$ of the proper time of the particle are related as $d\tau=dt/\gamma$, where $\gamma$ is the Lorentz factor of the particle. The four-force $F^i$ acting on the particle is given by
\begin{equation*}
\label{fourForce}
F^i=\frac{e}{c}F^{ik}\dot{x}_k,
\end{equation*}
where $F_{ik}=\partial A_k/\partial x^i-\partial A_i/\partial x^k$ is the electromagnetic-field tensor and the definition of the four-potential $A^i=(A^0,A^1,A^2,A^3)^T=(\phi,\mathbf{A}^T)^T$ contains the standard scalar $\phi$ and vector $\mathbf{A}$ electromagnetic potentials, with the electric and magnetic fields
given by the formulas
\begin{equation*}
\label{EMFieldsInTermsOfSimplePotentials}
\mathbf{E}=-\nabla\phi-\frac{1}{c}\frac{\partial\mathbf{A}}{\partial t},
\qquad\qquad
\mathbf{B}=\nabla\times\mathbf{A}.
\end{equation*}
The indices $i$ and $k$ take on the values $0,1,2,3$, with repeating indices everywhere denoting summation. The transition from contravariant to covariant components and vice versa is carried out using the metric tensor $g_{ik}=g^{ik}=\mathrm{diag}(1,-1,-1,-1)$.

We must obtain the equation of motion of the particle in the laboratory frame. This is straightforward using the following expressions for the derivatives of the four-vector $x^i$ with respect to the proper time of the particle:
\begin{equation}
\label{tauDerivatives}
\begin{aligned}
\dot{x}^i&=\gamma\frac{dx^i}{dt},\\
\ddot{x}^i&=\gamma\frac{d\gamma}{dt}\frac{dx^i}{dt}+\gamma^2\frac{d^2x^i}{dt^2},\\
\dddot{x}^i&=\gamma\Bigl(\frac{d\gamma}{dt}\Bigr)^2\frac{dx^i}{dt}+
\gamma^2\frac{d^2\gamma}{dt^2}\frac{dx^i}{dt}
+3\gamma^2\frac{d\gamma}{dt}\frac{d^2x^i}{dt^2}+\gamma^3\frac{d^3x^i}{dt^3},
\end{aligned}
\end{equation}

Let us make a transition to a set of dimensionless variables. We will measure the strengths of the electric and magnetic fields in units of the so-called critical
field $B_{cr}=m_e^2c^3/e\hbar\approx4.4\times10^{13}$~G, the particle velocity in units of the speed of light $c$, the particle charge in units of the positron charge $e$, the particle mass in units of the electron mass $m_e$, the particle energy in units of the electron rest energy $m_e c^2$, all distances in units of the Compton wavelength of the
electron $^-\!\!\!\!\lambda=\hbar/m_e c\approx3.86\times10^{-11}$~cm,
and all times in units of $^-\!\!\!\!\lambda/c$. Note that, in these units, $1000\text{ km}\approx2.6\times10^{18}$ and $1\text{ s}\approx7.8\times10^{20}$. The convenience of using these units is due to two factors: the presence of strong electromagnetic fields in the magnetosphere, which makes it reasonable to measure the field strengths in terms of the critical field, and the ultra-relativistic motion of the particles, which makes it natural to characterize the particle energy in terms of its Lorentz factor.

Using \eqref{tauDerivatives} after introduction of the above dimensionless variables and separation of the scalar and vector components of the four-vector $x^i$, the Dirac-Lorentz equation \eqref{DiracLorentzEquation} reduces to the system of equations
\begin{alignat}{2}
\label{dimentionlessDLEquation1}
\frac{d\gamma}{dt}&=\frac{2}{3}\alpha\gamma\left[\frac{d^2\gamma}{dt^2}
-\gamma^3\Bigl(\frac{d\mathbf{v}}{dt}\Bigr)^2\right]&&\pm\mathbf{v}\cdot\mathbf{E},
\\
\label{dimentionlessDLEquation2}
\gamma\frac{d\mathbf{v}}{dt}&=\frac{2}{3}\alpha\gamma\left[3\frac{d\gamma}{dt}
\frac{d\mathbf{v}}{dt}+\gamma\frac{d^2\mathbf{v}}{dt^2}\right]
&&\pm\biggl[\mathbf{E}-\mathbf{v}(\mathbf{v}\cdot\mathbf{E})+\mathbf{v}\times\mathbf{B}\biggr],
\end{alignat}
where $\mathbf{v}$ is the particle velocity and $\alpha=e^2/\hbar c\approx1/137$ --- is the fine-structure constant. When considering the motion of positrons, we should take a ``$+$'' sign in the system \eqref{dimentionlessDLEquation1} and
\eqref{dimentionlessDLEquation2}, and should take a ``$-$'' for the motion of electrons. For convenience, we will consider the ``$+$'' case, and comment when necessary on the changes that result if electrons rather than positrons are considered.

Equation \eqref{dimentionlessDLEquation1} represents the conservation of energy,
and \eqref{dimentionlessDLEquation2} the equation of motion of the particle.
This is easy to see if we formally specify the fine-structure constant to be $\alpha=0$ in \eqref{dimentionlessDLEquation1} and \eqref{dimentionlessDLEquation2}. These equations then take on their standard form, corresponding to neglecting the inverse influence of the field of the moving charged particle on the particle itself. The first of these equations indicates that the energy acquired by the particle per unit time is equal to the work done by the electric field on the particle. The second of these equations, \eqref{dimentionlessDLEquation2}, simply reflects the fact that the change in the relativistic momentum of the particle, $\mathbf{p}=\gamma\mathbf{v}$, per unit time is due to the action of the total Lorentz force, $\mathbf{E}+\mathbf{v}\times\mathbf{B}$.

One might ask why we are using the classical Dirac-Lorentz equation to describe particles in the electromagnetic field of the vacuum magnetosphere of a neutron star, and not the usual equations of motion. The reason is that the electric field is so strong that electrons and positrons are virtually instantaneously accelerated to relativistic energies, and there is intense emission of so-called curvature radiation when these particles move along the curved magnetic-field lines, due to their very high Lorentz factors, leading to energy losses by the particles. This radiative friction force is taken into account by the Dirac–Lorentz equation. The use of the usual equations of motion would lead to appreciable over-estimation of the particle energies, and therefore also of the energies of the photons emitted by the particles. In studies of amplification processes in the electron–positron plasma in the vacuum magnetosphere, this would lead to results that bear no relation to reality. The presence of radiative friction must be taken into account when numerically computing the trajectories of charged particles in the inner magnetosphere of a neutron star \citep{FinkbeinerEtal1989}. It is interesting that, in vacuum fields, this is also important at distances of the order of the light cylinder \citep{ZachariadesJackson1989,Zachariades1991}, when considering the behavior of charged particles in a wave field and a constant electric field \citep{Jackson1984}.

Let us consider this question in more detail. A charged particle created in the magnetosphere, be it an electron or positron, will experience an electrical force and be accelerated. Let us estimate the time required for the particle’s motion to become relativistic. For this, it is sufficient to use the equation of motion of the particle projected onto the direction of the magnetic field, $dp_\parallel/dt=E_\parallel$. Since the electric field does not change significantly over the time the particle is accelerated, we can immediately obtain the characteristic time $\tau_{rel}$ for the particle to make a transition to the relativistic regime, when the longitudinal momentum becomes close to $p_\parallel\simeq1$:
\begin{equation}
\label{tauRel}
\tau_{rel}\simeq\frac{1}{E_\parallel}.
\end{equation}
We see from the general form of \eqref{deutschMagneticField} and \eqref{deutschElectricField} how the electric field $\mathbf{E}$ is related to the magnetic field $\mathbf{B}$:
\begin{equation*}
\label{EinTermsOfB}
E\simeq \frac{R^2}{R_L r}B,
\end{equation*}
where the radius of the light cylinder in ordinary, dimensional units is $R_L=c/\Omega$, and takes on values for typical pulsar periods of $P\sim0.1{-}1$~s of $R_L\sim10^4{-}10^5$~km (we have used the fact that the wave number is $k=1/R_L$). We can find the electric field at the stellar surface $E_{surf}$ by setting $r=R$:
\begin{equation}
\label{Esurf}
E_{surf}\simeq \frac{R}{R_L}B_{surf}.
\end{equation}

For a typical surface magnetic field $B_{surf}\sim0.01{-}0.1$ and a ratio of the radii of the neutron star and the light cylinder $R/R_L\sim10^{-4}{-}10^{-3}$, the surface electric field is $E_{surf}\sim10^{-6}{-}10^{-4}$. Accordingly, the time for the transition to the relativistic regime is $\tau_{rel}\sim10^{4}{-}10^{6}$. This means that the particle will reach near-light speeds after a time of the order of $10^{-17}-10^{-15}$~s, after which its motion can be taken to be ultra-relativistic. Obviously, the particle will traverse a distance of no more than $10^{4}{-}10^{6}$ Compton wavelengths during this time. We have obtained an upper limit---the particle velocity has not reached the speed of light $c$ in the initial stage of the acceleration in the time interval considered. This justifies our assumption that the electric field does not change significantly during the particle’s acceleration time, since the acceleration time is small compared to the period of rotation of the neutron star, and the distance over which the acceleration occurs is small compared to the distance over which the electric field changes appreciably; in our case, this is a distance of the order of the radius of the star, $R\simeq10$~km.

Further, the particle will continue to be accelerated in an ultra-relativistic regime. As is well known, in general, the particle will move along the curved magnetic-field lines. Acquiring more and more energy, the particle begins to emit curvature radiation, whose characteristic energy $\varepsilon_{curv}$ and total intensity $W_{curv}$ are equal to
\begin{equation}
\label{curvatureRadiationCharacteristicEnergyAndTotalIntensity}
\varepsilon_{curv}=\frac{3}{2}\frac{\gamma^3}{\rho},\qquad\qquad W_{curv}=\frac{2}{3}\alpha\frac{\gamma^4}{\rho^2},
\end{equation}
where $\rho$ is the radius of curvature of the particle trajectory,
the energy $\varepsilon_{curv}$ is measured in units of $m_ec^2$ (like the particle energy), and the intensity $W_{curv}$ is measured in units of $m_ec^3/^-\!\!\!\!\lambda$. Since the intensity of curvature radiation grows with the Lorentz factor $\gamma$, the particle will eventually not undergo further acceleration after it has reached some maximum Lorentz factor $\gamma_0$, since all the energy it acquires from the electric field will be lost to curvature radiation. To determine $\gamma_0$, we must find a stationary solution to \eqref{dimentionlessDLEquation1}, having substituted in this equation $d\gamma/dt=d^2\gamma/dt^2=0$:
\begin{equation}
\label{stationaryECE}
\frac{2}{3}\alpha\gamma_0^4\Bigl(\frac{d\mathbf{v}}{dt}\Bigr)^2=\mathbf{v}\cdot\mathbf{E}.
\end{equation}
Equation \eqref{stationaryECE} shows that, in the stationary state, all the work done by the electric field on the particle is completely transformed into the energy of curvature radiation \eqref{curvatureRadiationCharacteristicEnergyAndTotalIntensity}. In the case considered, the particle’s velocity vector has unit length, $v=1$, so that $d\mathbf{v}/{dt}=\mathbf{n}/\rho$, where $\mathbf{n}$ is the principle normal vector to the particle trajectory and $\rho$ is the radius of curvature of the trajectory. The maximum Lorentz factor of the particle takes the form
\begin{equation}
\label{gammaMax}
\gamma_0=\left(\frac{3}{2\alpha}E_\parallel\rho^2\right)^{1/4}.
\end{equation}
For a characteristic longitudinal electric field $E_\parallel\sim10^{-4}$ and radius of curvature of the particle trajectory $\rho\sim R\simeq2.6\times10^{16}$, the maximum Lorentz factor of the particle is of the order of $\gamma_0\sim6\times10^7$. Here, we
have taken as a characteristic radius of curvature of the particle trajectory the radius $R$ of the neutron star, having in mind that the particle moves along trajectories close to the magnetic-field lines, and the radius of curvature of these lines is of the order of $R$ near the magnetic equator.

Let us estimate the time for a particle to acquire its maximum Lorentz factor $\gamma_0$---the time for the full acceleration of the particle and its transition to a quasi-stationary motion regime determined by the equilibrium condition \eqref{stationaryECE}. We use here the conservation of energy of the particle in its simplest form, $d\gamma/dt=E_\parallel$, without including the radiative-friction force. The desired time is then
\begin{equation}
\label{tauSt}
\tau_{st}\simeq\frac{\gamma_0}{E_\parallel}.
\end{equation}
We can see from \eqref{tauRel} and \eqref{tauSt} that $\tau_{st}=\gamma_0\tau_{rel}$; i.e., the time for the acquisition of the maximum Lorentz factor is a factor of $\gamma_0$ greater than the time for the particle to achieve near-light speed. Thus, the particle acquires its maximum energy over a time of the order of $\tau_{st}\sim10^{12}$ (in dimensional units, $10^{-9}$~s), after which the work by the electric field on the particle per unit time is equal to the total intensity of the curvature radiation. During this time, the particle traverses a distance of the order of $10^{12}$ (in dimensional units, several tens of centimeters), appreciably less than $R$; therefore, the assumption that the electric field does not change significantly over the total acceleration time $\tau_{st}$ is valid. This means that we can take the acceleration of the electrons and positrons to be virtually instantaneous, and to occur at the point where the electron–positron pair was created.

However, both the electric field strength and the radius of curvature of the trajectory change during the motion of the particle, so that the Lorentz factor $\gamma_0$ varies with time. This means that the particle will adjust its motion, acquiring energy due to work by the electric field if $\gamma_0$ grows along its trajectory or losing
energy to curvature radiation if $\gamma_0$ decreases along its trajectory. It is now important to determine the rate at which this adjustment of the particle’s energy occurs. If this rate is appreciably higher than the rate of variation of $\gamma_0$ along the trajectory, we can take the Lorentz factor to be determined by the coordinates above the point where the particle is located; at a fixed moment in time, the value of $\gamma_0$ depends only on the coordinates of the point considered, not on the velocity of the particle.

Let us find the variation of the Lorentz factor $\gamma$ of the particle as it approaches the steady-state value $\gamma_0$. Here, we use \eqref{dimentionlessDLEquation1}, having represented the Lorentz factor of the particle as the sum of the steady-state value $\gamma_0$ and some deviations from this value $\delta\gamma$. The deviations $\delta\gamma$ can be taken to be small, $\delta\gamma\ll\gamma_0$. This is true because we are considering how the Lorentz factor of the particle changes during its motion as a result of the smooth variation of $\gamma_0$ due to the variation of the longitudinal electric field $E_\parallel$ and the radius of curvature of the trajectory $\rho$, with the particle initially being in a stationary state and possessing an energy $\gamma_0$.
Using the smallness of the deviations $\delta\gamma$ to linearize \eqref{dimentionlessDLEquation1}, we obtain
\begin{equation}
\label{linearizedECL}
\frac{d^2\delta\gamma}{dt^2}-\frac{3}{2\alpha\gamma_0}\frac{d\delta\gamma}{dt}
-\frac{4\gamma_0^2}{\rho^2}\delta\gamma=0.
\end{equation}
The characteristic equation $\sigma^2-3\sigma/2\alpha\gamma_0-4\gamma_0^2/\rho^2=0$ for \eqref{linearizedECL} has the solution
\begin{equation}
\label{eigenValues}
\sigma_1=\frac{3}{2\alpha\gamma_0},\qquad
\sigma_2=-\frac{8}{3}\alpha\frac{\gamma_0^3}{\rho^2}.
\end{equation}
When computing these equations, we used the condition $|\sigma_2/\sigma_1|\ll1$. It follows from \eqref{gammaMax} and \eqref{eigenValues} that the condition that the ratio of the numbers themselves be small is equivalent to the condition
\begin{equation}
\label{lambdaCondition}
\frac{8}{3}\alpha E_\parallel\ll1.
\end{equation}

The condition \eqref{lambdaCondition} is always satisfied, and the indicated quantity of the order of $10^{-6}$ for the characteristic
longitudinal electric field, $E_\parallel\sim10^{-4}$. Note that the inequality \eqref{lambdaCondition} does not represent a constraint, and is satisfied even for the critical electric field $E_\parallel=1$ (it is well known that the electric-field strength cannot exceed this value due to the direct creation of electron–positron pairs from the vacuum in the presence of such strong fields).

The general solution of \eqref{linearizedECL} takes the form
\begin{equation}
\label{deltaGammaGeneralSolution}
\delta\gamma=C_1e^{\sigma_1 t}+C_2e^{\sigma_2 t}.
\end{equation}
As we can see from \eqref{eigenValues}, the eigenvalue $\sigma_1$ is positive. Formally, this would mean that a particle given some additional energy above $\gamma_0$ would begin to accelerate further, acquiring more and more energy. Clearly, this cannot happen physically. This is the so-called paradox of self-acceleration of the particle, well known in the theory of the Dirac–Lorentz equation \citep{SokolovTernov1974}. It arises because the Lorentz–Dirac equation \eqref{DiracLorentzEquation} contains a third derivative of the four-vector $x^i$ with respect to the proper time of the particle $\tau$. This means that it is not sufficient to know the initial coordinates and velocity of the particle in order to describe its motion, so that boundary conditions must be specified for this equation. These boundary conditions must be chosen to eliminate the self-accelerating solution. Apart from specifying the initial coordinates and velocity, we must require that the acceleration of the particle becomes zero after all external forces cease to act on it. In our case, this corresponds to having the acceleration of the particle vanish as $t\rightarrow\infty$. For the general expression \eqref{deltaGammaGeneralSolution}, this condition will be satisfied if and only if $C_1=0$, so that the self-accelerating solution disappears.

Taking this into account, we can immediately conclude that, when the Lorentz factor of the particle deviates by an amount $\delta\gamma_i$ from the stationary value $\gamma_0$, the Lorentz factor will approach $\gamma_0$ according to the exponential law
\begin{equation*}
\label{gammaEvolution}
\delta\gamma=\delta\gamma_i\,e^{-t/\tau_0}
\end{equation*}
with the decay constant
\begin{equation}
\label{tauA}
\tau_0=\frac{3}{8\alpha}\frac{\rho^2}{\gamma_0^3}.
\end{equation}
For characteristic radii of curvature of the trajectory $\rho\sim10^{17}$ and Lorentz factors $\gamma_0\sim10^7{-}10^8$, the decay time is $\tau_0\sim10^{11}{-}10^{14}$ (in dimensional units, $10^{-10}{-}10^{-7}$~s). Over a time $\tau_0$, the particle travels a
distance $l_0=\tau_0$ of the order of $10^{11}{-}10^{14}$ Compton wavelengths (i.e., from centimeters to several tens of meters). Now recall that the characteristic distances over which the electric field and radius of curvature of the magnetic-field lines---and therefore also the stationary Lorentz factor $\gamma_0$---vary are of the order of $R$. Because $l_0\ll R$, particles have time to adjust to variations in $\gamma_0$ during their motion in the magnetosphere. Thus, we can take the Lorentz factor of a particle to be determined fully by its coordinates.

This assertion requires some refinement. We have already discussed the fact that there exists a force-free surface \eqref{FFSequation} in the magnetosphere. It is clear that charged particles will move toward this force-free surface. However, the longitudinal electric field will gradually decrease as a particle approaches this surface. Simultaneously, the time for the adjustment of $\tau_0$ will increase. Eventually, the longitudinal electric field becomes sufficiently weak that the reaction time of the particle is too great to satisfy the condition of quasi-stationary motion; i.e., to support a balance between the work done by the electric field and the intensity of the curvature radiation. This is associated with the fact that the particle Lorentz factor decreases as the field weakens, together with the energy and intensity of curvature photons. As a consequence, the characteristic time for the particle energy losses near the force-free surface begins to exceed the characteristic time for variation of the electric field. We will consider this question in detail in our analysis of the capture of particles by the force-free surface.

Thus far, we have discussed the energetics of particles using only \eqref{dimentionlessDLEquation1}. Let us now consider in more detail the second equation \eqref{dimentionlessDLEquation2}, and investigate how this equation affects a particle’s trajectory. We will estimate the magnitudes of terms arising due to our allowance for the self-interaction of the charged particles, assuming that a time $\tau_{st}$ \eqref{tauSt} has passed after their creation, and that the particles have by this time already fully accelerated and made a transition into a quasi-stationary regime determined by the condition \eqref{stationaryECE}. Since the particle energy is determined by the
Lorentz factor $\gamma_0$ \eqref{gammaMax}, which varies over distances of the order of the radius of the neutron star $R$, $d\gamma/dt\sim\gamma_0/R$. The acceleration of the particle is of the order of $|d\mathbf{v}/dt|\sim1/R$, since the radius of curvature of the particle’s trajectory is close to the radius of curvature of the magnetic-field lines, which, in turn, is of the order of $R$, if we consider a region near the magnetic equator of the star. According to the same reasoning, the second time derivative of the particle velocity will be $|d^2\mathbf{v}/dt^2|\sim1/R^2$. Thus, the first term in \eqref{dimentionlessDLEquation2} containing the square brackets is of the order of $\alpha\gamma_0^2/R^2$. This is small compared to the second term, since
\begin{equation}
\label{selfActionNeglect1}
\alpha\frac{\gamma_0^2}{R^2}\ll E,
\end{equation}
as we can see by taking typical values of the Lorentz factor $\gamma_0\sim10^8$, the stellar radius $R\sim10^{17}$, and the electric field $E\sim10^{-4}$. However, the fulfilment of the condition \eqref{selfActionNeglect1} is not a sufficient basis to neglect terms allowing for the self-interaction. The term on the left-hand side of \eqref{dimentionlessDLEquation2} is to order of magnitude $\gamma_0/R\sim10^{-9}$, and is also appreciably smaller than $E$. To justify neglecting terms allowing for the influence of the fields created by the charged particles on their own motion, these terms must be small compared to $\gamma_0/R$:
\begin{equation*}
\label{selfActionNeglect2}
\alpha\frac{\gamma_0}{R}\ll1.
\end{equation*}
We see that this condition will always be satisfied, so that the motion of the particles in the vacuum magnetosphere of the neutron star will be described by the equation
\begin{equation}
\label{EqOfMotion}
\gamma\frac{d\mathbf{v}}{dt}=
\mathbf{E}-\mathbf{v}(\mathbf{v}\cdot\mathbf{E})+\mathbf{v}\times\mathbf{B}.
\end{equation}
This equation has been written for positrons; the corresponding equation for electrons will differ from \eqref{EqOfMotion} only in the presence of a minus sign before the entire right-hand side. Thus, Eq.~\eqref{EqOfMotion} describes the proper motion of the particle, while its energetics are determined by \eqref{dimentionlessDLEquation1}, which reduces to \eqref{gammaMax} in the quasi-stationary case.

Multiplying \eqref{EqOfMotion} vectorially by $\mathbf{B}$, we obtain an iterative formula for $\mathbf{v}_\perp$---the velocity component orthogonal to the magnetic field:
\begin{equation}
\label{vPerp}
\mathbf{v}_\perp=\frac{1}{B^2}\left[\mathbf{E}-\mathbf{v}(\mathbf{v}\cdot\mathbf{E})
-\gamma\frac{d\mathbf{v}}{dt}\right]\times\mathbf{B}.
\end{equation}

Generally speaking, charged particles in the strong magnetic field of a neutron star will have virtually no momentum orthogonal to this field: due to synchrotron cooling, the particles immediately go to the zeroth Landau level. This means that, in a first approximation, we can assume that the particle velocity is directed along the magnetic field, $\mathbf{v}=\mathbf{b}$, where $\mathbf{b}=\mathbf{B}/B$ is a unit tangent vector
to the local magnetic-field line. In a slightly more refined approximation, we must take into account the presence of a drift-velocity component orthogonal to the magnetic field. We will search for solutions in the form $\mathbf{v}=\mathbf{b}+\mathbf{v}_\perp$. Note that the quantity $\mathbf{v}_\perp$ is first order in $R/R_L$. After substituting the relation $\mathbf{v}=\mathbf{b}$ into \eqref{vPerp}, the total velocity of the particle $\mathbf{v}$ can be written
\begin{equation*}
\label{vFull}
\mathbf{v}=\mathbf{b}+\mathbf{v}_{e}+\mathbf{v}_{c},
\end{equation*}
where $\mathbf{v}_e$ is the electric drift and $\mathbf{v}_{c}$ the centrifugal drift
velocity. These velocities are defined by the formulas
\begin{align}
\label{electricDrift}
\mathbf{v}_{e}&=\frac{\mathbf{E}\times\mathbf{B}}{B^2},
\\
\label{anotherDrift}
\mathbf{v}_{c}&=\frac{\gamma}{B}\;\mathbf{b}\times\frac{d\mathbf{b}}{dt}.
\end{align}
It is easy to write the total time derivative of the tangent vector to the local magnetic-field line as
\begin{equation*}
\label{bDerivative}
\frac{d\mathbf{b}}{dt}=\frac{\partial\mathbf{b}}{\partial t}+(\mathbf{v}\cdot\nabla)\mathbf{b}.
\end{equation*}

The first term on the right-hand side is to order of magnitude $|\partial\mathbf{b}/\partial t|\sim\Omega$, or equivalently $\sim1/P$, and,
for typical pulsar periods $P\sim0.1{-}1$~s takes values $10^{-21}{-}10^{-20}$. Due to the condition $\mathbf{v}=\mathbf{b}$, the second term is $(\mathbf{b}\cdot\nabla)\mathbf{b}=\mathbf{n}/\rho$, where $\mathbf{n}$ is the principle
normal vector to the magnetic-field line and $\rho$ the radius of curvature of the field line. For typical radii of curvature $\rho\sim10^{17}$, the first term can be neglected compared to the second, so that the velocity \eqref{anotherDrift} mainly describes the centrifugal drift, which can be written
\begin{equation}
\label{centrifugalDrift}
\mathbf{v}_{c}=\frac{\gamma}{B\rho}\,\mathbf{e},
\end{equation}
where $\mathbf{e}=\mathbf{b}\times\mathbf{n}$ is the binormal vector. Note the difference in the directions of the centrifugal drift velocities for the electrons and positrons: formula \eqref{centrifugalDrift} is written for positrons, and for electrons, we must add a negative sign before the right-hand side. The electric drift velocity $\mathbf{v}_e$ for both electrons and positrons is given by \eqref{electricDrift}, and the direction of this velocity does not depend on the sign of the charged particle.

Let us estimate the drift velocities $\mathbf{v}_e$ and $\mathbf{v}_c$ to order of magnitude. The electric drift velocity is equal to the ratio of the electric and magnetic fields, $v_e\sim E/B$, and is of order $R/R_L\sim10^{-4}$ [see \eqref{Esurf}]. For $\gamma\sim10^8$, $\rho\sim10^{17}$, and $B\sim0.01{-}0.1$, the centrifugal drift velocity is $v_c\sim10^{-8}{-}10^{-7}$. Thus, the centrifugal drift velocity is the next order of smallness compared to the electric drift velocity, and we will accordingly neglect the centrifugal drift. We will assume that
\begin{equation}
\label{finalFullVelocity}
\mathbf{v}=\mathbf{b}+\mathbf{v}_e
\end{equation}
with accuracy to within terms $o(kr)$, i.e., to within quantities of quadratic and higher order in $kr\sim R/R_L$. Note that, for \eqref{finalFullVelocity}, the equality $v=1$ is satisfied in this case with accuracy to first order in $R/R_L$, and a discrepancy appears only in second order. If it were necessary for us to take into account quantities of the order of $\sim10^{-8}$, we would have to calculate the following terms in the expansion of order $(R/R_L)^2$ by iterating \eqref{vPerp}. This would lead to corrections to the longitudinal velocity of the particle, and it would no longer be possible to assume $\mathbf{v}_\parallel\approx\mathbf{b}$. The satisfaction of \eqref{finalFullVelocity} and the smallness of the drift for the orthogonal motion compared to the longitudinal motion enabled us to assume that the particles essentially move along the magnetic-field lines.

The electromagnetic field described by \eqref{deutschMagneticField} and \eqref{deutschElectricField} is periodic in time. This leads to a certain inconvenience when considering the motion of a particle in the laboratory frame. If the particle travels some distance $dl$ between points 1 and 2 over a time $dt$, the variation of the electromagnetic field at the final point 2 is determined not only by the variation in the coordinates, but also by the fact that the field itself at the point 2 varies over the time $dt$. Let $(t,r,\theta,\varphi)$ be the spherical coordinates in the laboratory frame. We transform to the new coordinates $(t',r',\vartheta',\varphi')$ using the relations $t'=t$, $r'=r$, $\vartheta'=\theta$, $\varphi'=\varphi-\Omega t$. This leads to the transformation of the partial derivatives
$\partial/\partial t=\partial/\partial t'-\Omega\,\partial/\partial\varphi'$,
$\partial/\partial r=\partial/\partial r'$, $\partial/\partial\theta=\partial/\partial\vartheta'$,
$\partial/\partial\varphi=\partial/\partial\varphi'$. We will call $(t',r',\vartheta',\varphi')$ the coordinates in the rotating frame. In these coordinates, the electromagnetic field depends only on $(r',\vartheta',\varphi')$, not on $t'$. Further, we will omit the primes for variables when it is clear that we are considering quantities in the rotating frame. The transformation of the velocity and acceleration in the transition to the rotating system have the standard form
\begin{equation}
\label{velocityAccelerationTransform}
\begin{aligned}
\mathbf{v}&=\mathbf{v}'+\mathbf{v}_{tr},\\
\frac{d\mathbf{v}}{dt}&=\frac{d\mathbf{v}'}{dt'}
+2\,\mathbf{\Omega}\times\mathbf{v}'+\mathbf{\Omega}\times\mathbf{v}_{tr},
\end{aligned}
\end{equation}
where $\mathbf{v}'$ is the relative velocity and $\mathbf{v}_{tr}=\mathbf{\Omega}\times\mathbf{r}$ the translational velocity. Formulas \eqref{velocityAccelerationTransform} are valid when $\dot{\Omega}\ll\Omega^2$.
This is equivalent to the condition that $\dot{P}\ll1$, which is always satisfied, since, to order of magnitude, $\dot{P}\sim10^{-15}$. Substituting \eqref{velocityAccelerationTransform} into \eqref{EqOfMotion} yields for the equation of motion of the particle in the rotating frame
\begin{equation}
\label{EqOfMotionInRotatingFrame}
\gamma\frac{d\mathbf{v}'}{dt'}=
\mathbf{E}+\mathbf{v}_{tr}\times\mathbf{B}
-\mathbf{v}'(\mathbf{v}'\cdot\mathbf{E})+\mathbf{v}'\times\mathbf{B}.
\end{equation}
Here, $\mathbf{E}$ and $\mathbf{B}$ are the electric and magnetic fields, which are determined by \eqref{deutschElectricField} and \eqref{deutschMagneticField}, but depend on the coordinates $(r',\vartheta',\varphi')$ after the described change of variables.

When obtaining \eqref{EqOfMotionInRotatingFrame}, we neglected the Coriolis, $2\,\mathbf{\Omega}\times\mathbf{v}'$, and translational, $\mathbf{\Omega}\times\mathbf{v}_{tr}$, accelerations compared to the relative acceleration $d\mathbf{v}'/dt'$. This is possible because the relative acceleration is primarily axipetal, and due to the motion of the particles along the curved magnetic-field lines. Therefore, $|d\mathbf{v}'/dt'|\sim1/R$, which is to order of magnitude $10^{-17}$. To order of magnitude, the Coriolis acceleration is $|2\,\mathbf{\Omega}\times\mathbf{v}'|\sim1/R_L$, which is $10^{-21}$. The translational acceleration is still smaller, because the translational speed $v_{tr}$ is clearly less than $v'\sim1$ in the considered regions in the magnetosphere, $r\lesssim10R\ll R_L$, where the efficient single-photon creation of pairs is possible. Moreover, only when $r\ll R_L$ are \eqref{deutschMagneticField} and \eqref{deutschElectricField} for the electromagnetic field valid. In general, in this case, the ratio of the translational to the Coriolis acceleration is equal to the ratio of the Coriolis to the relative acceleration, and is $R/R_L\sim10^{-4}$. Thus, the neglected term in the left-hand side of \eqref{EqOfMotionInRotatingFrame}, $\gamma\,\mathbf{\Omega}\times(2\mathbf{v}'+\mathbf{v}_{tr})$, is also small compared to the electric-field strength, $E\sim10^{-4}$; for $\gamma\sim10^8$, it is to order of magnitude $\gamma/R_L\sim10^{-13}$. In the right-hand side of \eqref{EqOfMotionInRotatingFrame} are left only terms whose magnitudes are comparable to $E$, while all substantially smaller terms are neglected. The term $\mathbf{v}'\times\mathbf{B}$ is retained, because, although $v'\sim1$, the velocity vector itself $\mathbf{v}'$ is nearly parallel to the magnetic field, and its orthogonal component $\mathbf{v}'_\perp$, which is the only one to contribute to the vector product, is to order of magnitude $v_e\sim E/B$, so that $\mathbf{v}'\times\mathbf{B}$ is also of the order of $E$. The Lorentz factor $\gamma$ in \eqref{EqOfMotionInRotatingFrame} is measured in the laboratory frame.

Further, we will investigate the motion of the particles in the rotating frame. We are interested in whether there exist regions in the magnetosphere where the accumulation of primary plasma is possible. We will find equilibrium positions---points where a charged particle can remain for an indefinitely long time. The coordinates of the equilibrium positions are determined by the conditions that the velocity and acceleration be equal to zero in the rotating frame: $\mathbf{v}'=d\mathbf{v}'/dt'=0$. Substituting this condition into \eqref{EqOfMotionInRotatingFrame} leads to the equation
\begin{equation}
\label{EqForEquiPoints}
\mathbf{E}^{eff}=0,
\end{equation}
were we have introduced the effective electric field $\mathbf{E}^{eff}=\mathbf{E}+\mathbf{v}_{tr}\times\mathbf{B}$. The components of this field, which can straightforwardly be found from \eqref{deutschMagneticField},
\eqref{deutschElectricField} together with the expression for $\mathbf{v}_{tr}$, have the following form in spherical coordinates
\begin{equation}
\label{effectiveElectricField}
\begin{aligned}
E^{eff}_r&=-kr\left[\frac{R^2}{r^2}\cos\theta B_r+\Bigl(1-\frac{R^2}{r^2}\Bigr)
\sin\theta B_\theta\right],\\
E^{eff}_\theta&=kr\Bigl(1-\frac{R^2}{r^2}\Bigr)\left[\frac{1}{2}\sin\theta B_r+\cos\theta B_\theta\right],\\
E^{eff}_\varphi&=kr\Bigl(1-\frac{R^2}{r^2}\Bigr)\cos\theta B_\varphi,
\end{aligned}
\end{equation}
where $B_r$, $B_\theta$ and $B_\varphi$ are determined by \eqref{deutschMagneticField}.
Substituting the condition \eqref{EqForEquiPoints} into \eqref{effectiveElectricField} yields the set of equilibrium points
\begin{equation}
\label{nonisolatedEquiPoints}
\begin{alignedat}{2}
r&=R,&\quad\theta&=\frac{\pi}{2}\qquad\text{(equator)},\\
r&=R,&\quad\theta'&=\frac{\pi}{2}\qquad\text{(magnetic equator)},
\end{alignedat}
\end{equation}
\begin{equation}
\label{isolatedEquiPointsAtLeafs}
\begin{aligned}
\frac{r_{+}^2}{R^2}&=\frac{3+\cos\theta_m}{1-\cos\theta_m}
\qquad\text{(open sheets)},
\\
(\theta,\varphi)&=\left\{\Bigl(\frac{\theta_m}{2},\varphi_m\Bigr),
\Bigl(\pi-\frac{\theta_m}{2},\pi+\varphi_m\Bigr)\right\},
\end{aligned}
\end{equation}
\begin{equation}
\label{isolatedEquiPointsAtDomes}
\begin{aligned}
\frac{r_{-}^2}{R^2}&=\frac{3-\cos\theta_m}{1+\cos\theta_m}
\qquad\text{(folds)},
\\
(\theta,\varphi)&=\left\{\Bigl(\frac{\pi}{2}+\frac{\theta_m}{2},\varphi_m\Bigr),
\Bigl(\frac{\pi}{2}-\frac{\theta_m}{2},\pi+\varphi_m\Bigr)\right\}.
\end{aligned}
\end{equation}

We can see that all the equilibrium points are located on the force-free surface \eqref{FFSequation} and can be divided into two groups: non-isolated and isolated. The nonisolated equilibrium points are all on the equator and magnetic equator [see \eqref{nonisolatedEquiPoints}]. Two isolated equilibrium points are added to these regions: two whose coordinates are given by \eqref{isolatedEquiPointsAtLeafs}, located on open sheets of the force-free surface, and two others given by \eqref{isolatedEquiPointsAtDomes} on the folding parts of the surface adjacent to its boundary with the equator and magnetic equator. All four of these points lie in the plane passing through the rotational axis and magnetic axis. Only the points \eqref{isolatedEquiPointsAtLeafs} lie in the cross section of the bisectrix angle
between $\mathbf{\Omega}$ and $\mathbf{m}$ with the force-free surface, while the points \eqref{isolatedEquiPointsAtDomes} lie on the cross section of the line orthogonal to this bisectrix lying in the plane of the vectors $\mathbf{\Omega}$ and $\mathbf{m}$ with the same force-free surface. The radial coordinates $r_{-}$ of the points lying on cupolas are always smaller than the radial coordinates $r_{+}$ of the points lying on open sheets of the force-free surface. The coordinates $r_{+}$ and $r_{-}$ are equal, $r_{+}=r_{-}=\sqrt{3}R$, only if $\theta_m=\pi/2$. If $\theta_m\rightarrow0$, then $r_{-}\rightarrow R$ and $r_{+}\rightarrow\infty$; the equality $r_{+}/R=2\sqrt{2}/\theta_m$ is asymptotically obeyed for small values of $\theta_m$.

This raises the question of the stability of the equilibrium positions we have found. Moreover, the character of the trajectories of the particle motion is not clear, both near the force-free surface in general and near the equilibrium points in particular. We will provide an answer to this question after a detailed study of the capture of particles by the force-free surface.

\section{OSCILLATIONS OF CHARGED PARTICLES NEAR THE FORCE-FREE SURFACE}

Let us consider the motion of charged particles near the force-free surface, applying some qualitative reasoning. We can see immediately that the motion will be oscillatory, because the longitudinal electric field changes its sign in the transition through the force-free surface. Here, we will assume that the signs of the charged particles are such that the electric force on them is directed toward the force-free surface. Further, some regular motion along the force-free surface will be superposed on the oscillatory motion of a charged particle. If we imagine carrying out an average over the rapid oscillations, thereby distinguishing the leading center, or the local equilibrium position about which the particle oscillates, this center can only be located on the force-free surface. Otherwise, on average, a non-zero electric field would act on the particle, causing it to return to the surface.

We will study the particle motion quantitatively in the rotating frame using \eqref{EqOfMotionInRotatingFrame}. We choose some point $\mathbf{r}_0$ on the force-free surface and expand the electric field $\mathbf{E}$, magnetic field $\mathbf{B}$, and effective electric field $\mathbf{E}^{eff}$ introduced above about this point as follows:
\begin{equation}
\label{fieldExpansion}
\begin{aligned}
\mathbf{E}&=\mathbf{E}_0+(\mathbf{x'}\cdot\nabla)\mathbf{E}_0,\\
\mathbf{B}&=\mathbf{B}_0+(\mathbf{x'}\cdot\nabla)\mathbf{B}_0,\\
\mathbf{E}^{eff}&=\mathbf{E}^{eff}_0+(\mathbf{x'}\cdot\nabla)\mathbf{E}^{eff}_0,
\end{aligned}
\end{equation}
where $\mathbf{E}_0$, $\mathbf{B}_0$, and $\mathbf{E}^{eff}_0$ are the field strengths at the point $\mathbf{r}_0$, and $\mathbf{x'}=\mathbf{r}-\mathbf{r}_0$ is the distance from
the point $\mathbf{r}$ for which we are interested in the field strengths to the point $\mathbf{r}_0$. The expansion \eqref{fieldExpansion} is valid because we are considering the motion of a particle in the immediate vicinity of $\mathbf{r}_0$, at distances appreciably smaller than the characteristic distances for variations of the fields, so that $x'\ll R$. Before searching for the full solution of \eqref{EqOfMotionInRotatingFrame}, we will first find some partial solution, describing the motion of a particle with constant velocity $\mathbf{v}'_0=\mathrm{const}$. This solution will then satisfy the system of equations
\begin{equation}
\label{particularSolutionSystem}
\begin{aligned}
\frac{d\mathbf{x}'_0}{dt'}&=\mathbf{v}'_0,\\
0&=\mathbf{E}^{eff}_0+(\mathbf{x}'_0\cdot\nabla)\mathbf{E}^{eff}_0
+\mathbf{v}'_0\times\mathbf{B}_0+\mathbf{v}'_0\times(\mathbf{x}'_0\cdot\nabla)\mathbf{B}_0.
\end{aligned}
\end{equation}
When obtaining the second equation, we used the fact that $d\mathbf{v}'_0/dt'=0$. We also required that $v'_0\ll1$, and neglected the term $\mathbf{v}'_0(\mathbf{v}'_0\cdot\mathbf{E})$, which is small compared to $\mathbf{E}$, and therefore also compared to $\mathbf{E}^{eff}$. It follows from the first equation of \eqref{particularSolutionSystem} that $\mathbf{x}'_0=\mathbf{v}'_0 t$ (at the initial time, the particle is located at the point $\mathbf{r}_0$ on the force-free surface). Substituting $\mathbf{x}'_0$ into the second equation yields
\begin{equation*}
\label{transformedSecondEq}
\mathbf{E}^{eff}_0+\mathbf{v}'_0\times\mathbf{B}_0
+t\Bigl[(\mathbf{v}'_0\cdot\nabla)\mathbf{E}^{eff}_0
+\mathbf{v}'_0\times(\mathbf{v}'_0\cdot\nabla)\mathbf{B}_0\Bigr]
=0.
\end{equation*}
For the motion of the particle to occur with the same constant velocity at the next moment in time, two conditions must be satisfied:
\begin{equation}
\label{conditionsForPartSolution}
\begin{aligned}
\mathbf{E}^{eff}_0+\mathbf{v}'_0\times\mathbf{B}_0=0,\\
(\mathbf{v}'_0\cdot\nabla)\mathbf{E}^{eff}_0
+\mathbf{v}'_0\times(\mathbf{v}'_0\cdot\nabla)\mathbf{B}_0=0.
\end{aligned}
\end{equation}

The first equation of \eqref{conditionsForPartSolution} unambiguously determines the velocity component orthogonal to the magnetic field
\begin{equation}
\label{fullDrift}
\mathbf{v}'_\perp=\frac{\mathbf{E}_0^{eff}\times\mathbf{B}_0}{B_0^2},
\end{equation}
with the longitudinal component $v'_\parallel$ remaining arbitrary:
\begin{equation}
\label{partSolVelocity}
\mathbf{v}'_0=v'_\parallel\mathbf{b}+\mathbf{v}'_\perp.
\end{equation}
The form of \eqref{fullDrift} for the velocity $\mathbf{v}'_\perp$ coincides with that of expression \eqref{electricDrift} for the electric drift velocity in the laboratory frame, with $\mathbf{E}$ replaced by $\mathbf{E}^{eff}$. It is easy to verify using \eqref{fullDrift} that the drift velocity $\mathbf{v}'_\perp$ in the rotating frame is indeed given by the sum of the electric drift velocity in the laboratory frame and the ``translational'' drift, equal to $\mathbf{b}(\mathbf{v}_{tr}\cdot\mathbf{b})-\mathbf{v}_{tr}=-{\mathbf{v}_{tr}}_\perp$, which is simply the negative of the component of the translational velocity orthogonal to the magnetic field. This is true because we are considering the motion in the rotating frame; the presence of the translational velocity $\mathbf{v}_{tr}=\mathbf{\Omega}\times\mathbf{r}$ provides an additional contribution to the drift, so that, in the laboratory frame, $\mathbf{v}_e=\mathbf{v}'_\perp+{\mathbf{v}_{tr}}_\perp$. We introduce the velocity components $\mathbf{v}'_\perp$:
\begin{equation*}
\label{driftInRotatingFrame}
\begin{aligned}
v'_{\perp r}&=
\Omega r\sin\theta\,b_r b_\varphi\frac{1}{2}\,\Bigl(1-\frac{R^2}{r^2}\Bigr),\\
v'_{\perp \theta}&=
\Omega r\!\left[\cos\theta\,b_r+\Bigl(1-\frac{R^2}{r^2}\Bigr)\sin\theta\,
b_\theta\right]b_\varphi,\\
v'_{\perp \varphi}&=
-\Omega r\!\left[\Bigl(1-\frac{R^2}{r^2}\Bigr)\sin\theta\,\Bigl(\frac{1}{2}b_r^2+b_\theta^2\Bigr)
+\cos\theta\,b_r b_\theta\right],
\end{aligned}
\end{equation*}
where $b_r$, $b_\theta$, and $b_\varphi$ are components of the unit vector $\mathbf{b}$.

The longitudinal component $v'_\parallel$ of the velocity $\mathbf{v}'_0$ is unambiguously determined by the second equation of \eqref{conditionsForPartSolution}. Taking the scalar product of this equation and the vector $\mathbf{B}_0$ yields
\begin{equation*}
\label{multEq}
\mathbf{B}_0\cdot(\mathbf{v}'_0\cdot\nabla)\mathbf{E}^{eff}_0
+(\mathbf{B}_0\times\mathbf{v}'_0)\cdot(\mathbf{v}'_0\cdot\nabla)\mathbf{B}_0=0.
\end{equation*}
Using the fact that $\mathbf{E}^{eff}_0=-\mathbf{v}'_0\times\mathbf{B}_0$, we immediately obtain
\begin{equation}
\label{ortogonalityCondition}
\mathbf{v}'_0\cdot\nabla(\mathbf{E}_0\cdot\mathbf{B}_0)=0.
\end{equation}
Here, we have also used the fact that $\mathbf{E}^{eff}_0\cdot\mathbf{B}_0=\mathbf{E}_0\cdot\mathbf{B}_0$. However, since the gradient $\nabla(\mathbf{E}_0\cdot\mathbf{B}_0)$ is directed normal to the force-free surface, $\mathbf{E}\cdot\mathbf{B}=0$, the velocity $\mathbf{v}'_0$ lies in the tangent plane passing through the point $\mathbf{r}_0$ of the force-free surface.

Thus, the partial solution describing the motion of a particle with constant velocity exists and is given by \eqref{partSolVelocity}. The longitudinal velocity component is obtained from the condition \eqref{ortogonalityCondition}, and has the form
\begin{equation*}
\label{longitudinalVelocity}
v'_\parallel=-\frac{\mathbf{v}'_\perp\cdot\nabla(\mathbf{E}_0\cdot\mathbf{B}_0)}
{\mathbf{b}\cdot\nabla(\mathbf{E}_0\cdot\mathbf{B}_0)}.
\end{equation*}

We can see that the velocity $\mathbf{v}'_0$ lies in the tangent plane, so that the particle cannot leave the force-free surface, and its trajectory lies entirely on this surface. The velocity will vary from point to point, and is determined by the same expression \eqref{partSolVelocity}, but with the fields at the point where the particle is located at the given time used in place of $\mathbf{E}_0^{eff}$ and $\mathbf{B}_0$.

However, the solution found does not exhaust all classes of motion of the particle. This is clear, if for no other reason than because a solution with velocity $\mathbf{v}'_0$ describes adiabatic motion of the particle along the force-free surface, with no oscillations of the particle whatsoever. We will now find the full solution of \eqref{EqOfMotionInRotatingFrame} near the force-free surface. We seek a solution in the
form
\begin{equation}
\label{solutionForm}
\mathbf{x}'=\mathbf{x}'_0+\mathbf{x}'_1,\qquad
\mathbf{v}'=\mathbf{v}'_0+\mathbf{v}'_1,
\end{equation}
where $\mathbf{x}'_0$ and $\mathbf{v}'_0$ are the adiabatic solution found above. No special restrictions are applied to the quantities $\mathbf{x}'_1$ and $\mathbf{v}'_1$ a priori, apart from the requirement that $\mathbf{x}'_1$ be small compared to the characteristic scale for variation of the electromagnetic field, $R$. In particular, the velocity $\mathbf{v}'_1$ can be close to the speed of light. Substituting \eqref{solutionForm} into \eqref{EqOfMotionInRotatingFrame} and using \eqref{particularSolutionSystem} and \eqref{fieldExpansion} leads to the equation
\begin{equation}
\label{EqOfMotionNearFFS}
\gamma\frac{d\mathbf{v}'_1}{dt'}=
(\mathbf{x}'_1\cdot\nabla)\mathbf{E}^{eff}_0+\mathbf{v}'_0\times(\mathbf{x}'_1\cdot\nabla)\mathbf{B}_0
+\mathbf{v}'_1\times\mathbf{B}-\mathbf{v}'_1(\mathbf{v}'_1\cdot\mathbf{E}),
\end{equation}
where $\mathbf{E}$ and $\mathbf{B}$ are determined by \eqref{fieldExpansion} and the coordinate value $\mathbf{x}'$ by \eqref{solutionForm}. When deriving \eqref{EqOfMotionNearFFS}, we used the condition $v'_0\ll1$. The term $\mathbf{v}'_1(\mathbf{v}'_1\cdot\mathbf{E})$ is kept because, generally speaking, the velocity $\mathbf{v}'_1$ can be of the order of unity, so that this term can be comparable to $\mathbf{E}$.

Let us first consider the non-relativistic case ($v'_1\ll1$). It immediately follows from \eqref{EqOfMotionNearFFS} that
\begin{equation}
\label{nonrelativisticOscillationEq}
\frac{d\mathbf{v}'_1}{dt'}=
(\mathbf{x}'_1\cdot\nabla)\mathbf{E}^{eff}_0+\mathbf{v}'_0\times(\mathbf{x}'_1\cdot\nabla)\mathbf{B}_0.
\end{equation}
Here, we have used the fact that $\mathbf{v}'_1\approx v'_1\mathbf{b}$, i.e., the velocity is directed essentially along the magnetic field, so that the term $\mathbf{v}'_1\times\mathbf{B}$ vanishes. It is obvious that the linear differential equation \eqref{nonrelativisticOscillationEq} describes oscillations. The left-hand side contains the second time derivative of the vector $\mathbf{x}'_1$, and the form of the right-hand side is linear in the components of the vector $\mathbf{x}'_1$, and can be represented as the product of a $3\times3$ matrix and the vector $\mathbf{x}'_1$. However, we can convince ourselves of the oscillatory character of the solution without searching for the eigenvalues of this matrix and corresponding eigenvectors. Let us find the frequency for the non-relativistic oscillations. We take the scalar product of \eqref{nonrelativisticOscillationEq} with the vector $\mathbf{B}_0$. Carrying out manipulations on the right-hand side of the resulting equation analogous to those described above for the derivation of \eqref{ortogonalityCondition} straightforwardly yields for the square of the frequency of the non-relativistic oscillations
\begin{equation}
\label{nonrelativisticOscillationsFrequency}
\omega^2=-\frac{\mathbf{b}\cdot\nabla(\mathbf{E}_0\cdot\mathbf{B}_0)}{\mathbf{B}_0}.
\end{equation}
In this case, $\mathbf{x}'_1\approx x'_1\mathbf{b}$ and $\mathbf{v}'_1\approx v'_1\mathbf{b}$, and the coordinate $x'_1$ satisfies the oscillation equation $d^2x'_1/dt^2=-\omega^2x'_1$. When deriving \eqref{nonrelativisticOscillationsFrequency}, we also assumed implicitly that the instantaneous direction of the vector $\mathbf{b}$ at the point where the particle is located coincides with the direction of the vector $\mathbf{b}_0$ at the point $\mathbf{r}_0$ about which we expanded the fields, i.e., $\mathbf{b}_0\approx\mathbf{b}$.

Let us find criteria for applicability of the non-relativistic approximation. As we can see from \eqref{nonrelativisticOscillationsFrequency}, the characteristic oscillation frequency is $\omega\sim\sqrt{\Omega B}$, or equivalently,
\begin{equation}
\label{characteristicFrequency}
\omega\sim\sqrt{\frac{B}{R_L}}.
\end{equation}
Here, we have used the estimate \eqref{Esurf} for the electric field at the surface of the neutron star. For characteristic light-cylinder radii $R_L\sim10^{20}$ (for pulsars with $P\sim1$~s) and magnetic fields $B\sim0.01{-}0.1$ the non-relativistic oscillation frequency is of order $\omega\sim10^{-11}{-}10^{-10}$ (which corresponds to $\nu=\omega/2\pi\sim1{-}10\text{ GHz}$ in dimensional units). In the
non-relativistic case, the $v'_1\ll1$, and the oscillating particle travels a distance that is clearly less than $1/\nu$, during one period. The maximum amplitude of the non-relativistic oscillations is
\begin{equation}
\label{maxNonrelativisticAmpolitude}
l_{nro}\simeq\frac{1}{\omega}.
\end{equation}
The oscillatory motion will be non-relativistic if $x'_1\ll l_{nro}$. For the frequencies $\omega$ found above, we have $l_{nro}\sim10^{10}{-}10^{11}$ ($0.1{-}1$~cm in dimensional units). This means that the equality $\mathbf{b}_0=\mathbf{b}$ used when deriving \eqref{nonrelativisticOscillationsFrequency} is precise by virtue of the extreme smallness of $l_{nro}$ compared to $R$. We can see that oscillations of a charged particle about the force-free surface with an amplitude exceeding $1$~cm are clearly relativistic. Thus, our primary interest is in considering the relativistic case of oscillations, which is realized in nature.

To obtain an equation describing the particle oscillations without placing any special constraints on the velocity, we must use \eqref{EqOfMotionNearFFS}, setting in this equation $\mathbf{v}'_1\parallel\mathbf{B}$:
\begin{equation}
\label{relativisticOscillationEq}
\gamma\frac{d\mathbf{v}'_1}{dt'}=
(\mathbf{x}'_1\cdot\nabla)\mathbf{E}^{eff}_0+\mathbf{v}'_0\times(\mathbf{x}'_1\cdot\nabla)\mathbf{B}_0
-\mathbf{v}'_1(\mathbf{v}'_1\cdot\mathbf{E}).
\end{equation}
Taking the scalar product of \eqref{relativisticOscillationEq} with $\mathbf{B}_0$ yields
\begin{equation}
\label{relativisticOscillationEq1}
\gamma\frac{dv'_1}{dt'}=-\omega^2x'_1-(v'_1)^2E_\parallel,
\end{equation}
where $\omega$ is the frequency of the non-relativistic oscillations given by \eqref{nonrelativisticOscillationsFrequency}. We can see immediately
that the term $(v'_1)^2E_\parallel$ becomes negligible and $\gamma\approx1$ when $v'_1\ll1$, so that \eqref{relativisticOscillationEq1} is transformed into the usual equation for non-relativistic oscillations with frequency $\omega$. Now let the velocity $v'_1$ not be small. We must then use the condition $x'_1\ll R$ and write the longitudinal electric field in the form $E_\parallel=-\omega^2x'_1$. After using the relation $\gamma\approx1/\sqrt{1-(v'_1)^2}$, we quickly obtain the equation for relativistic oscillations of a charged particle about the force-free surface:
\begin{equation}
\label{absRelativisticOscillationEq}
\frac{dv'_1}{dt'}=-\frac{\omega^2}{\gamma^3}x'_1.
\end{equation}

We will first make a number of qualitative comments concerning the ultra-relativistic regime for the particle’s motion, when its velocity is close to the velocity of light, $v'_1=1$, during a large fraction of the oscillation period. Let us find the regions where the particle’s motion is non-relativistic. It is clear that these regions are located near the turning points of the motion, where the particle’s velocity vanishes and its positional coordinate reaches the amplitude $A$. The width of the regions of non-relativistic motion near the turning points is then determined by the so-called acceleration length $l_a$, equal to the distance over which a particle that is initially at a turning point and has zero velocity acquires near-light speed. Another way of expressing this is to say that, after traveling a distance equal to the acceleration length $l_a$, the relativistic momentum of the particle $\mathbf{p}$ becomes comparable to unity. An upper limit for $l_a$ is given by the amplitude $l_{nro}$ for non-relativistic oscillations of the particle [see Eq.~\eqref{maxNonrelativisticAmpolitude}]. Indeed, if a particle that is initially at rest deviates by the distance $l_{nro}$ from the equilibrium position $\mathbf{x}'_0$ and attains relativistic energies when it passes through the force-free surface, a particle located at a turning point will clearly be accelerated over a shorter time. This is true because the electric field at the turning point is non-zero, and exceeds the nearly-zero field in the region of non-relativistic oscillations. A lower limit for $l_a$ is given by the characteristic time for the acceleration of particles to relativistic velocities in the magnetosphere $\tau_{rel}$ (see \eqref{tauRel}). Thus, $\tau_{rel}\ll l_a\ll l_{nro}$. It is easy to obtain a more accurate estimate of the acceleration length $l_a$ near a turning point using the same reasoning as was used when deriving \eqref{tauRel}, but setting $E_\parallel\sim\omega^2A$:
\begin{equation*}
\label{accelerationLength}
l_a\simeq\frac{l^2_{nro}}{A}.
\end{equation*}
The ultra-relativistic approximation is applicable if $l_a\ll A$. Note that the acceleration length $l_a$ is smaller than $l_{nro}$ by a factor of $A/l_{nro}$, i.e., appreciably less than 1~cm. This obviates the need to consider the exact character of the motion of a charged particle near its turning points for oscillation amplitudes of the order of several $l_{nro}$. In this case, the particle moves from one turning point to the other with a velocity $v'_1=1$, so that the oscillation period is $T=4A$. The oscillations themselves are given approximately by
\begin{equation}
\label{ultrarelativisticOscillationsFormula}
x'_1=\frac{2A}{\pi}\arcsin\sin\Bigl(\frac{\pi t}{2A}\Bigr)
\end{equation}
and represent saw-like oscillations with amplitude $A$, with $x'_1=t$ when $-A\leqslant t\leqslant A$ (further, we do not distinguish $t$ and $t'$, since $t=t'$).

Nevertheless, \eqref{absRelativisticOscillationEq} can be analyzed exactly without using the ultra-relativistic approximation. This equation has the first integral
\begin{equation*}
\label{firstIntegral}
C=\gamma+\frac{\omega^2(x'_1)^2}{2}.
\end{equation*}
This first integral is simply equal to the maximum Lorentz factor $\gamma_{\max}$ reached by the charged particle in its motion through the force-free surface; i.e., $C=\gamma_{\max}$ when $x'_1=0$. The first integral $C$ can be expressed in terms of the oscillation amplitude:
\begin{equation}
\label{cThroughA}
C=1+\frac{\omega^2A^2}{2}.
\end{equation}

It can be shown that, after the following changes of variables,
\begin{equation}
\label{changeOfVariables}
\begin{alignedat}{2}
&\sin{\phi_0}=\frac{\kappa}{a},&\qquad\sin{\phi}&=\frac{\omega}{\sqrt{2C}}\frac{x'_1}{a},\\
&\kappa=\sqrt{\frac{C-1}{C+1}},&\qquad a&=\sqrt{\frac{C-1}{C}},
\end{alignedat}
\end{equation}
Eq. \eqref{absRelativisticOscillationEq} can be written in the differential form
\begin{equation*}
\label{EqOfRelativisticOscillationsInDiff}
dt=\pm\frac{\sqrt{2C}}{\omega}\frac{1}{\sin\phi_0}\,d[E(\phi,\kappa)-\cos^2\!\phi_0F(\phi,\kappa)].
\end{equation*}
Here, $F(\phi,\kappa)$ and $E(\phi,\kappa)$ are elliptical integrals of the first and second kind, respectively, defined as is done by Gradshteyn and Ryzhik \citep{GradsteinRyzhik1963}:
\begin{equation}
\label{ellipticIntegrals}
\begin{aligned}
F(\phi,\kappa)&=\int\limits_0^\phi\frac{d\alpha'}{\sqrt{1-\kappa^2\sin^2\alpha'}},\\
E(\phi,\kappa)&=\int\limits_0^\phi\sqrt{1-\kappa^2\sin^2\alpha'}\:d\alpha'.
\end{aligned}
\end{equation}
Introducing the function $R(\phi,\kappa)=E(\phi,\kappa)-\cos^2\!\phi_0F(\phi,\kappa)$, we obtain the relation between $\phi$ and $t$
\begin{equation}
\label{phitRelation}
t=\frac{\sqrt{2C}}{\omega}\frac{R(\phi,\kappa)}{\sin\phi_0}.
\end{equation}
We can immediately exactly determine the oscillation frequency, noting that $\phi=0$ corresponds to the equilibrium position and $\phi=\pi/2$ to a turning point. The distance from the equilibrium position $x'_1=0$ to the turning point $x'_1=A$ is traversed by the particle over one-quarter of its period, so that
\begin{equation}
\label{accurateT}
T=4\frac{\sqrt{2C}}{\omega}\frac{\mathbf{R}(\kappa)}{\sin\phi_0},
\end{equation}
where $\mathbf{R}(\kappa)=\mathbf{E}(\kappa)-\cos^2\!\phi_0\mathbf{K}(\kappa)$, and
$\mathbf{K}(\kappa)=F(\pi/2,\kappa)$ and $\mathbf{E}(\kappa)=E(\pi/2,\kappa)$ are total elliptical integrals of the first and second kinds.

Let us verify the asymptotic oscillation period \eqref{accurateT}. If the oscillations are non-relativistic ($A\ll l_{nro}$), as we can see from \eqref{cThroughA} and \eqref{changeOfVariables}, $C\simeq1$,
$\kappa\simeq0$, $\sin\phi_0\simeq1/\sqrt{2}$, and $\mathbf{K}(0)=\mathbf{E}(0)=\pi/2$, so that $T=2\pi/\omega$ is the usual period for non-relativistic oscillations. If the oscillations are ultra-relativistic ($A\gg l_{nro}$), then $C\simeq\omega^2A^2/2$, $\kappa\simeq\sin\phi_0\simeq1$, and $\mathbf{E}(1)=1$, so that $T=4A$. We obtained this same result above based on qualitative reasoning. Note that, in spite of the logarithmic divergence of $\mathbf{K}(\kappa)$ as $\kappa\rightarrow1$, the expression $\cos^2\!\phi_0\mathbf{K}(\kappa)$ vanishes as $\kappa\rightarrow1$, because it then goes as $(\kappa'^2/2)\ln(4/\kappa')$, where we have introduced the notation $\kappa'=\sqrt{1-\kappa^2}$.

The exact solution of the oscillation equation \eqref{absRelativisticOscillationEq} can be written in the form
\begin{equation}
\label{oscSolution}
x'_1=A\sin\phi,
\end{equation}
where
\begin{equation}
\label{phiExpr}
\phi=\mathcal{Q}\biggl(\frac{\omega t}{\sqrt{2(C+1)}}\biggr),
\end{equation}
as follows from \eqref{changeOfVariables} and \eqref{phitRelation}. Here, we have introduced the function $\mathcal{Q}(z)$, inverse to the function $R(\phi,\kappa)$, such that $z=R(\mathcal{Q}(z),\kappa)$ for any real number $z\in\mathbb{R}$. This is possible because $R(\phi,\kappa)$ grows strictly and is continuously differentiable with respect to the variable $\phi$ on the entire real $\mathbb{R}$ axis when $0\leqslant\kappa<1$, with $R(\mathbb{R},\kappa)=\mathbb{R}$. When these conditions are satisfied, the inverse function $\mathcal{Q}(z)$ exists and is also a single-valued, strictly growing, continuously differentiable function on $\mathbb{R}$, so that $\mathcal{Q}(\mathbb{R})=\mathbb{R}$. The derivative $d\mathcal{Q}/dz$ is never equal to zero or infinity, since this is true for the partial derivative $\partial R(\phi,\kappa)/\partial\phi$ for all $C$ values of interest to us in the interval $1\leqslant C<\infty$ (this corresponds to any physically possible value of the particle Lorentz factor $\gamma_{\max}$).

Formulas \eqref{oscSolution} and \eqref{phiExpr} exhaust the problem of oscillations of a charged particle about the force-free surface in the absence of radiative energy losses. We do not require such an analysis of the function $\mathcal{Q}(z)$; let us note just one of its properties. As follows from the analogous properties of the elliptical integrals \eqref{ellipticIntegrals}, the function $R(\phi,k)$ satisfies the relation
\begin{equation*}
\label{RquasiperiodicRelation}
R(\pi n\pm\phi,k)=2n\mathbf{R}(k)\pm R(\phi,k)
\end{equation*}
for an arbitrary whole number $n\in\mathbb{Z}$. Thanks to this, it is true for the function $\mathcal{Q}(z)$ that
\begin{equation*}
\label{QquasiperiodicRelation}
\mathcal{Q}(z+2n\mathbf{R}(k))=\mathcal{Q}(z)+\pi n.
\end{equation*}
If we pass from $\mathcal{Q}(z)$ to $\phi(t)$ using \eqref{phiExpr}, the equality $\phi(t+nT/2)=\phi(t)+\pi n$ follows for $\phi(t)$, where $T$ is determined by \eqref{accurateT}. Hence, we can again convince ourselves that a charge particle will indeed undergo oscillations with the period $T$. For this, it is sufficient to take any even $n$ and substitute the resulting relation for $\phi(t)$ into \eqref{oscSolution}.

Asymptotically, the functions $\mathcal{Q}(z)$ for $\kappa$ close to zero and to unity have the form
\begin{equation}
\label{qAsymptotics}
\mathcal{Q}(z)=
\begin{cases}
\begin{alignedat}{2}
&2\,z,&\qquad \kappa&=0,\\
&\arcsin (z-2h(z))+\pi h(z),&\qquad \kappa&\rightarrow1,
\end{alignedat}
\end{cases}
\end{equation}
where the integral values of the function $h(z)$ are determined by the formula
\begin{equation*}
\label{hz}
h(z)=\left\lfloor\frac{z+1}{2}\right\rfloor.
\end{equation*}
Here, $\lfloor y\rfloor$ denotes the integer part of the real number $y$. No constraints are imposed on $z$ in \eqref{qAsymptotics}.

In the case of non-relativistic oscillations,we must use the asymptotic of the function $\mathcal{Q}(z)$ when $\kappa=0$. As we can see from \eqref{phiExpr}, $z=\omega t/2$ when $C\simeq1$. We then immediately obtain the equation $x'_1=A\sin(\omega t)$ for harmonic oscillations with frequency $\omega$. In the ultra-relativistic case, $C\gg1$, which corresponds to $\kappa\rightarrow1$, and substituting the corresponding asymptotic of $\mathcal{Q}(z)$ \eqref{qAsymptotics} into \eqref{oscSolution} yields
\begin{equation}
\label{relOscillations}
x'_1=A(-1)^{h(z)}\left(z-2h(z)\right),
\end{equation}
where, as follows from \eqref{phiExpr}, $z=t/A$. We can see that this expression is simply another form of \eqref{ultrarelativisticOscillationsFormula} obtained above based on qualitative reasoning. Relations \eqref{ultrarelativisticOscillationsFormula} and \eqref{relOscillations} are fully equivalent.

\section{CONCLUSION}

The dynamics of the motion of electrons and positrons in the inner vacuum magnetosphere of a neutron star can be represented as follows. A charged particle created far from the force-free surface will reach relativistic speeds over a time $\tau_{rel}\sim10^{-17}-10^{-15}$~s \eqref{tauRel}, and will make a transition to a quasi-stationary motion regime over a time $\tau_{st}\sim10^{-9}$~s \eqref{tauSt}, having traversed a distance of the order of $10-100$~cm. The Lorentz factor is then $\gamma_0\sim10^7-10^8$ \eqref{gammaMax}, and is fully determined by the balance between the power obtained from the accelerating electric field, $E_{\parallel}/B_{cr}\sim10^{-6}-10^{-4}$ \eqref{Esurf} (in dimensionless form), and the intensity of the curvature radiation. The particle moves essentially along the magnetic-field line \eqref{finalFullVelocity}, since the electric drift velocity is of order $v_e/c\sim10^{-4}$ \eqref{electricDrift} and the centrifugal drift velocity \eqref{anotherDrift}, \eqref{centrifugalDrift} is even smaller. The radius of curvature $\rho$ and longitudinal electric field $E_{\parallel}$ slowly change along the particle’s trajectory, leading to an adjustment in the particle Lorentz factor $\gamma_0$. The time for this readjustment is fairly small ($\tau_0\sim10^{-10}-10^{-7}$~s \eqref{tauA}), and the particle traverses a distance of order $1\text{ cm}-100\text{ m}$ over this time, whose upper limit is achieved near the force-free surface. This distance is appreciably less than the radius of the star, so that the Lorentz factor is essentially determined by coordinates of the particle. As a charged particle approaches the force-free surface, the quasi-stationary condition is disrupted. As a consequence, the particle passes through the force-free surface and begins to undergo adiabatic, ultra-relativistic oscillations. These oscillations decay due to radiative energy losses, while their frequency grows. When the amplitude $l_{nro}\sim1$~cm \eqref{maxNonrelativisticAmpolitude} is achieved, the oscillations become non-relativistic and harmonic, at the frequency $\nu\sim1-10$~GHz
\eqref{characteristicFrequency}. Since the regular component of the particle’s velocity lies in the plane tangent to the force-free surface, the particle undergoes a regular drift motion along the force-free surface, simultaneously with its oscillatory motion.

\section*{ACKNOWLEDGMENTS}

This work was partially supported by the Russian Foundation for Basic Research (project code 08-02-00749-a), the Program of State Support for Leading Scientific Schools of the Russian Federation (grant no. NSh-1738.2008.2) and the State
Agency for Science and Innovation (state contract no.~02.740.11.0250).
\newpage
\renewcommand{\refname}
\begin{flushright}
\textit{Translated by D. Gabuzda}
\end{flushright}
\newpage
\begin{figure}
\begin{center}
\includegraphics[height=0.89\textheight]{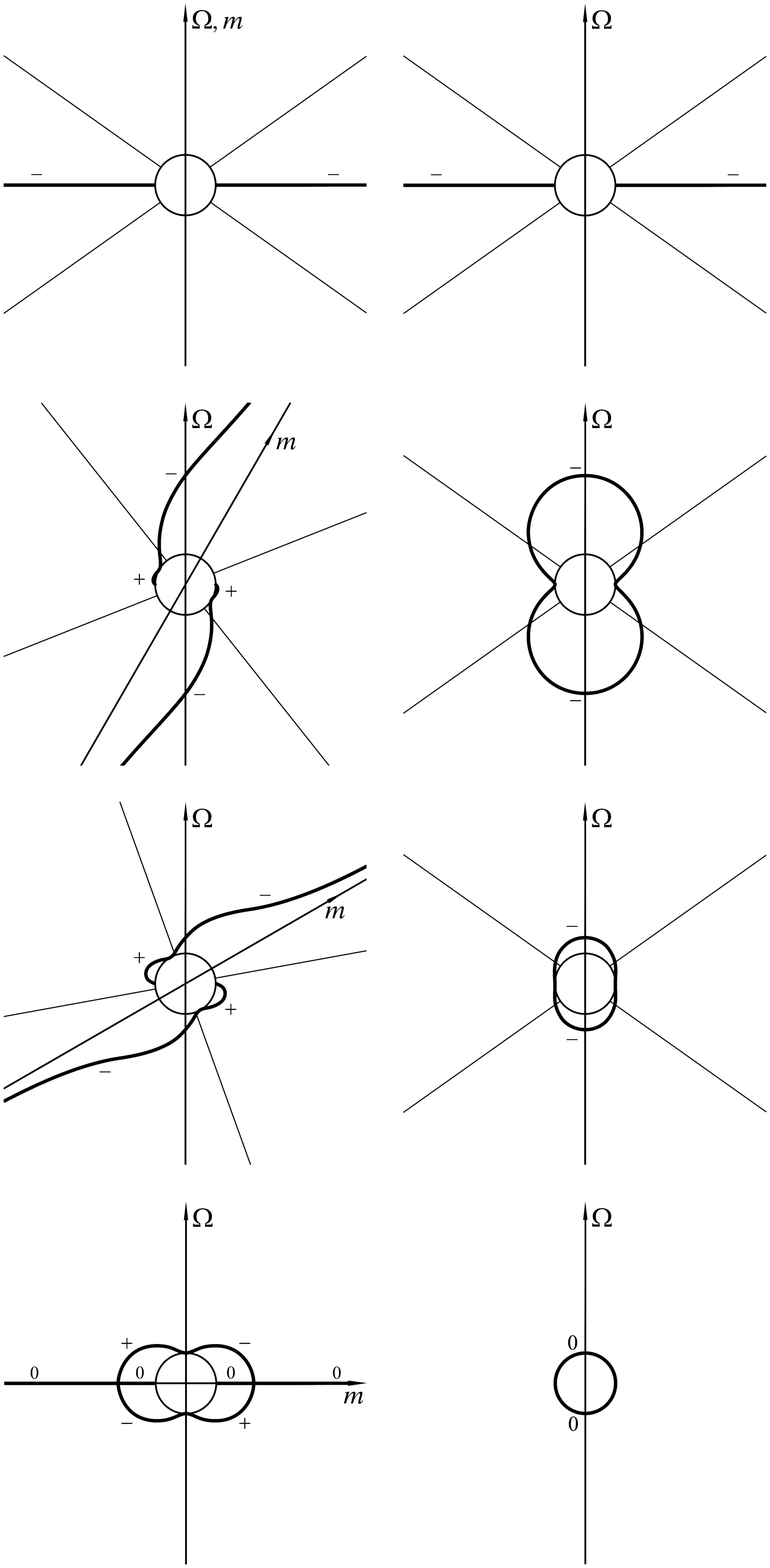} 
\end{center}
Cross sections of the force-free surface by the plane $\varphi-\varphi_m=\{0,\pi\}$ (left) and the plane $\varphi-\varphi_m=\{\pi/2,3\pi/2\}$ (right) for angles between the magnetic and rotational axes of $0$, $\pi/6$, $\pi/3$, $\pi/2$. (top to bottom). The signs of the accumulating charges are indicated. For comparison, we also show the cross section $\rho_{GJ}=0$.
\end{figure}

{\large\centering REFERENCES}
\begin{thebibliography}{99}
\bibitem[Sturrock(1971)]{Sturrock1971}
P. A. Sturrock,
Astrophys. J. \textbf{164}, 529 (1971).
\bibitem[Ruderman and Sutherland(1975)]{RudermanSutherland1975}
M. A. Ruderman and P. G. Sutherland,
Astrophys. J. \textbf{196}, 51 (1975).
\bibitem[Klepikov(1954)]{Klepikov1954}
N. P. Klepikov,
Zh. Eksp. Teor. Fiz. \textbf{26}, 19 (1954).
\bibitem[Erber(1966)]{Erber1966}
T. Erber,
Rev. Mod. Phys. \textbf{38}, 626 (1966).
\bibitem[Goldreich and Julian(1969)]{GoldreichJulian1969}
P. Goldreich and W. H. Julian,
Astrophys. J. \textbf{157}, 869 (1969).
\bibitem[Gurevich and Istomin(2007)]{GurevichIstomin2007}
A. V. Gurevich and Ya. N. Istomin,
Mon. Not. R. Astron. Soc. \textbf{377}, 1663 (2007).
\bibitem[Kramer et~al.(2006)]{KramerEtal2006}
M. Kramer, A. G. Lyne, J. T. O'Brien, et~al.,
Science \textbf{312}, 549 (2006).
\bibitem[Kramer(2008)]{Kramer2008}
M. Kramer,
AIP Conf. Proc. \textbf{983}, 11 (2008).
\bibitem[Ritchings(1976)]{Ritchings1976}
R. T. Ritchings,
Mon. Not. R. Astron. Soc. \textbf{176}, 249 (1976).
\bibitem[Wang et~al.(2007)]{WangEtal2007}
N. Wang, R. N. Manchester, and S. Johnston,
Mon. Not. R. Astron. Soc. \textbf{377}, 1383 (2007).
\bibitem[McLaughlin et~al.(2006)]{McLaughlinEtal2006}
M. A. McLaughlin, A. G. Lyne, D. R. Lorimer, et~al.,
Nature \textbf{439}, 817 (2006).
\bibitem[Zhu and Xu(2006)]{ZhuXu2006}
W. W. Zhu and R. X. Xu,
Mon. Not. R. Astron. Soc. \textbf{365}, L16 (2006).
\bibitem[Zhang et~al.(2007)]{ZhangEtal2007}
B. Zhang, J. Gil, and J. Dyks,
Mon. Not. R. Astron. Soc. \textbf{374}, 1103 (2007).
\bibitem[Li(2006)]{Li2006}
X.-D. Li,
Astrophys. J. \textbf{646}, L139 (2006).
\bibitem[Lomiashvili et~al.(2007)]{LomiashviliEtal2007}
D. Lomiashvili, G. Machabeli, and I. Malov,
arXiv:0709.2019 [astro-ph] (2007).
\bibitem[Ouyed et~al.(2008)]{OuyedEtal2008}
R. Ouyed, D. Leahy, B. Niebergal, et~al.,
arXiv:0802.3929 [astro-ph] (2008).
\bibitem[Esamdin et~al.(2008)]{EsamdinEtal2008}
A. Esamdin, C. S. Zhao, Y. Yan, et~al.,
Mon. Not. R. Astron. Soc. \textbf{389}, 1399 (2008).
\bibitem[Istomin and Soby'anin(2007)]{IstominSobyanin2007}
Ya. N. Istomin and D. N. Sobyanin,
Astron. Lett. \textbf{33}, 660 (2007).
\bibitem[Istomin and Sobyanin(2008)]{IstominSobyanin2008}
Ya. N. Istomin and D. N. Sobyanin,
AIP Conf. Proc. \textbf{983}, 298 (2008).
\bibitem[Deutsch(1955)]{Deutsch1955}
A. J. Deutsch,
Ann. d'Astrophys. \textbf{18}, 1 (1955).
\bibitem[Tiomno(1973)]{Tiomno1973}
J. Tiomno,
Phys. Rev. D \textbf{7}, 992 (1973).
\bibitem[Jackson(1978)]{Jackson1978}
E. A. Jackson,
Astrophys. J. \textbf{222}, 675 (1978).
\bibitem[Michel and Li(1999)]{MichelLi1999}
F. C. Michel and H. Li,
Phys. Rep. \textbf{318}, 227 (1999).
\bibitem[Soper(1972)]{Soper1972}
S. R. K. Soper,
Astrophys. Space Sci. \textbf{19}, 249 (1972).
\bibitem[Ferrari and Trussoni(1973)]{FerrariTrussoni1973}
A. Ferrari and E. Trussoni,
Astrophys. Space Sci. \textbf{24}, 3 (1973).
\bibitem[Finkbeiner et~al.(1989)]{FinkbeinerEtal1989}
B. Finkbeiner, H. Herold, T. Ertl, and H. Ruder,
Astron. Astrophys. \textbf{225}, 479 (1989).
\bibitem[Zachariades and Jackson(1989)]{ZachariadesJackson1989}
H. A. Zachariades and E. A. Jackson,
Phys. Rev. A \textbf{40}, 3769 (1989).
\bibitem[Zachariades(1991)]{Zachariades1991}
H. A. Zachariades,
Astrophys. Space Sci. \textbf{176}, 105 (1991).
\bibitem[Jackson(1984)]{Jackson1984}
E. A. Jackson,
J. Math. Phys. \textbf{25}, 1584 (1984).
\bibitem[Sokolov and Ternov(1974)]{SokolovTernov1974}
A. A. Sokolov and I. M. Ternov,
\textit{The Relativistic Electron} (Nauka, Moscow, 1974), p.~128 [in Russian].
\bibitem[Gradshteyn and Ryzhik(1963)]{GradsteinRyzhik1963}
I. S. Gradshteyn and I. M. Ryzhik,
\textit{Tables of Integrals, Series and Products} (Academic, New York, 2000; Moscow, GIFML, 1963), p.~918.
\end{thebibliography}
\end{document}